\documentclass{article}
\usepackage[ansinew]{inputenc}
\usepackage{latexsym}
\newcommand{\be}{\begin{equation}}
\newcommand{\ee}{\end{equation}}
\newcommand{\bear}{\begin{eqnarray}}
\newcommand{\ear}{\end{eqnarray}}
\newcommand{\ba}{\begin{eqnarray*}}
\newcommand{\ea}{\end{eqnarray*}}
\renewcommand{\theequation}{\arabic{section}.\arabic{equation}}

\newcommand{\dt}{\mbox{\boldmath$:$}}
\newcommand{\vdt}{\mbox{\bf$\vdots$}}

\newcommand{\no}{\noindent}

\newcommand{\lc}{\mbox{\Large$\ell$}}
\newcommand{\mZ}{{\mathsf{Z}}}

\newcommand{\z}{\mathit{z}}

\begin{document}

\title{Quantum Electrodynamics in Two-Dimensions at Finite Temperature\\Thermofield Bosonization Approach}
\author{L. V. Belvedere$^\ast$, R. L. P. G. Amaral$^\ast$, K. D. Rothe$^{\ast\ast}$ and A. F. Rodrigues$^{\ast\ast\ast}$\\
\small{$\ast$ Instituto de F\'{\i}sica}\\
\small{Universidade Federal Fluminense}\\
\small{Av. Litor\^anea S/N, Boa Viagem, Niter\'oi, CEP. 24210-340}\\
\small{Rio de Janeiro - Brasil}\\
\small{$\ast \ast $ Institut f\"ur Theoretische Physik}\\
\small{Universit\"at Heidelberg}\\
\small{Philosophenweg 16, D-69120}\\
\small{ Heidelberg - Germany}\\
\small{$\ast\ast\ast$  Centro Brasileiro de Pesquisas F\'{\i}sicas (CBPF)}\\
\small{Rua Dr. Xavier Sigaud, 150 - Urca - CEP. 22290-180}\\
\small{Rio de Janeiro - RJ - Brasil }\\
{\bf Latex file:QED2FT-submited}}
\date{\today}

\maketitle

\begin{abstract}
The Schwinger model at finite temperature is analyzed using the Thermofield Dynamics formalism. The operator solution  due to Lowenstein and Swieca is generalized to the case of finite temperature within the thermofield bosonization approach. The general properties of the statistical-mechanical ensemble averages of observables in the Hilbert subspace of gauge invariant thermal states are discussed. The  bare charge and chirality of the Fermi thermofields are screened, giving rise to an infinite number of mutually orthogonal thermal ground states. One consequence of the bare charge and chirality selection rule at finite temperature is that there are innumerably many thermal vacuum states with the same total charge and chirality of the doubled system. The fermion charge and chirality selection rules at finite temperature turn out to imply the existence of a family  of thermal theta vacua states parametrized with the same number of parameters as in zero temperature case. We compute the thermal theta-vacuum expectation value of the mass operator and show that the analytic expression of the chiral condensate for any temperature is easily obtained  within this approach, as well as, the corresponding high-temperature behavior.

\end{abstract}

\section{Introduction}

Quantum electrodynamics of massless fermions in two dimensions (Schwinger model) was exactly solved by Schwinger using functional methods \cite{Schwinger}. The original motivation was to show that local gauge invariance does not necessarily require the existence of a massless physical particle. However, it was only later that the subtleties which make this model particularly interesting were displayed by the work of Lowenstein and Swieca \cite{LS}. Paralleling Klaiber's paper on the Thirring model \cite{Klaiber}, a complete operator solution in terms of Bose fields (bosonization) was presented, and ``only then does the striking simplicity of the physical content of the model become obvious '' \cite{LS}. During four decades the Schwinger model has been extensively discussed within different approaches, applications and extensions. (For a general survey and a complete list of references see Refs. \cite{AAR,AARII}).

Using the functional integral approach within the imaginary time formalism, the extension of the Schwinger model to the case of finite temperature has been the subject of several publications \cite{AARII,Das,DJ,Love,Kao,Sachs,KaoLee}. The main conclusions are : $i)$ the temperature independence of the anomaly, which in turn is responsible for the temperature independence of the Schwinger mass; $ii)$ a lengthly computation of the temperature dependence of the chiral condensate shows that the spontaneous chiral symmetry breaking persists at finite temperature; $iii)$ the correlator of the Polyakov loop-operator in the ``zero instanton'' sector leads to clustering violation for any finite temperature. This suggests the existence of a degeneracy of the ground state, although Lorentz invariance is not manifest at finite temperature.

As is well known, some properties of a two-dimensional quantum field model are more transparent in the operator formulation, and others are better seen in the functional integral formulation. The purpose of the present paper is to analyze the operator formulation of the  Schwinger model with massless fermions at finite temperature within the thermofield dynamics approach, using the bosonization of Fermi fields at finite temperature (thermofield bosonization), which has been  introduced in Refs. \cite{ABR,ABRII}. From the present analysis some known results, formerly obtained within the functional integral formulation \cite{AARII,Das,DJ,Kao,Sachs,KaoLee}, are easily obtained within the thermofield bosonization approach. 

However, despite of a number of publications on  the Schwinger model at finite temperature, some questions related to the basic structural properties of the model  have not been fully appreciated and clarified in a convincing manner in the literature, such as : i) the role played by the bare fermionic charge and chirality selection rules in the construction of the Hilbert space of thermal physical states; ii) the charge screening in the Higgs-Schwinger mechanism at finite temperature, giving rise to an infinite number of thermal vacuum states; iii) the formulation of the gauge invariant Hilbert subspace in terms of a thermal theta vacuum parametrization, displaying the spontaneous  symmetry breaking at finite temperature.  It is the purpose of this paper to discuss these aspects of
the model, as well as to acquire a more detailed understanding of the physical grounds of gauge theories at finite temperature within the thermofield dynamics approach.   

The introduction of finite temperature first requires a doubling of the Hibert space at zero temperature in order to allow for a later  extension to $T\neq 0$. In the present case this requires a doubling of the Lowenstein and Swieca operator solution \cite{LS,RS}, which, as we shall
explain, involves some subtleties. This is done in Section 2. In order to have a clear understanding of the role played by the bare charge and chirality selection rules, a detailed discussion of the superselection rules and vacuum structure in the doubled Hilbert space at zero temperature is  presented.  

In Section 3 we discuss the operator solution of the Schwinger model at finite temperature. In thermofield dynamics the statistical-mechanical ensemble averages are expressed in the form of expectation values of $T=0$ operators in a temperature-dependent vacuum. In the case of a gauge theory, such statistical averages can for instance be  performed in the gauge invariant subspace defined by the Gupta-Bleuler subsidiary condition \cite{LS,O}. We show that at finite temperature the free fermionic charge and chirality of the total combined system are screened, giving rise to an infinite number of degenerate thermal vacua states. The thermal selection rules imply that a theta vacuum  representation in the Hilbert space of the gauge invariant thermal states can be given in terms  of a family of thermal theta vacua parametrized by the same number of  parameters as in the zero temperature case. 

In Section 4 we compute the thermal theta-vacuum expectation value of the mass operator and obtain the analytic expression for the chiral condensate for any temperature in terms of the mean number of massive particles in the ensemble. The high temperature behavior of the chiral condensate is then easily obtained within the thermofield bosonization approach. In Section 5 we conclude and present a brief discussion of other physical properties of the model. 

In the Appendix we consider the two-dimensional free massive scalar thermofield. The two-point functions are computed. We present a general construction of the off-diagonal thermal two-point function (for Bose and Fermi thermofields, massive or massless) as an analytic continuation of the diagonal ones, in agreement with Umezawa's view \cite{U}. This streamlines the presentation of Ref. \cite{ABR}. 

\section{QED$_2$ with doubled Hilbert space at $T = 0$}
\setcounter{equation}{0}

In thermofield dynamics a ``tilde'' operator is introduced for each of the operators describing the system under consideration \cite{Das,O,U,TFD,Mats}. This entails a doubling of the Hilbert space. In this section we shall reconsider the operator solution of the Schwinger model, as formulated by Lowenstein and Swieca \cite{LS}, taking account of this doubling of the Hilbert space. The temperature dependence of the expectation values (statistical-mechanical ensemble averages) of operators living in this doubled Hilbert space will then be entirely contained in  the thermal vacuum state to be discussed in the next section. This will set the stage for making the transition to finite temperature in section 3.

The doubled Schwinger model at zero temperature is defined by the classical total Lagrangian of the combined system ,

\[
\widehat{\cal L} (x)  = {\cal L} (x) - \widetilde{\cal L} (x)\,,
\]

\no where ${\cal L}$ is the Lagrangian of the original system
\footnote{The conventions used are: 

$$\gamma^0 = \pmatrix{0 & 1 \cr 1 & 0}\,,\,\gamma^1 
= \pmatrix{0 & 1 \cr - 1 & 0}\,,\,\gamma^5 = \gamma^0 \gamma^1\,\,,\,\,
\epsilon^{0 1} =  1\,,\,g^{00} = 1\,,
$$
$$
\gamma^\mu \gamma^5 = \epsilon^{\mu\nu} \gamma_\nu\,,\,x^\pm = x^0 \pm x^1\,,\,\partial_\pm 
= \partial_0 \pm \partial_1\,.
$$ },
\be\label{L}
{\cal L} (x) =\, - \,\frac{1}{4}\,{\cal F}_{\mu \nu} (x){\cal F}^{\mu \nu} (x)
+ \overline\psi (x) \gamma^\mu \Big ( i \partial_\mu + e {\cal A}_\mu (x) 
\Big )\psi (x)\,,
\ee

\no and ${\widetilde{\cal L}}$ is the Lagrangian of the ``tilde'' system obtained from (\ref{L}) by the ``tilde'' conjugation rule \cite{O},

\[
\widetilde{\cal L} (x) = \,-\, \frac{1}{4}\,\widetilde{\cal F}_{\mu \nu} (x)
\widetilde{\cal F}^{\mu \nu} (x)
+ \overline{\widetilde\psi} (x) \gamma^\mu \Big ( - i \partial_\mu + e 
\widetilde{\cal A}_\mu (x)  
\Big )\widetilde\psi (x)\,,
\]

\no with ${\cal F} (x)$ and $\widetilde{\cal F} (x)$ the field-strength tensors for the vector fields ${\cal A}_\mu(x)$  and $\widetilde{\cal A}\mu(x)$. At $T=0$ the contribution of the tilde- and untilde fields decouple in the correlation functions. The operator solution of the model will be given in terms of Wick-ordered exponentials of a set of free Bose fields and their corresponding ``tilde conjugated'' partners, acting on the Fock vacuum state

\be\label{Fock-vacuum}
\vert \widetilde 0 , 0 \rangle = \vert \widetilde 0 \rangle \otimes \vert 0 \rangle\,.
\ee

\no  At zero temperature, the corresponding quantum field theory is defined by the sets of field operators $\{\psi,{\cal A_\mu}\}$ and $\{\tilde\psi,\tilde{\cal A_\mu}\}$, which define two independent field algebras $\mathbf{A}$ and $\widetilde\mathbf{A}$, respectively, and generate two independent Hilbert spaces: 

\[
\widetilde{\cal H}\otimes {\cal H} = \widetilde{\mathbf{A}} \vert \widetilde 0 \rangle \otimes \mathbf{A} \vert 0 \rangle\,.
\]

\no In the local covariant operator formulation, the operator solutions for the Dirac equations  

\[
i \gamma^\mu \partial_\mu \psi (x)\,+\,\frac{e}{2}\, \gamma^\mu\,\lim_{{\varepsilon \rightarrow 0}\atop{\varepsilon^2 < 0}} \Big \{\,{\cal A}_\mu (x + \varepsilon ) \psi (x)\,+\,\psi (x)\,{\cal A}_\mu (x - \varepsilon )\,\Big \} = 0\,,
\]
\[
-\,i \gamma^\mu \partial_\mu \widetilde\psi (x)\,+\,\frac{e}{2}\, \gamma^\mu\,\lim_{{\varepsilon \rightarrow 0}\atop{\varepsilon^2 < 0}} \Big \{\,\widetilde{\cal A}_\mu (x + \varepsilon ) \widetilde\psi (x)\,+\,\widetilde\psi (x)\,\widetilde{\cal A}_\mu (x - \varepsilon )\,\Big \} = 0\,,
\]

\no are essentially obtained by the doubling of the Lowenstein and Swieca solution \cite{LS},
taking proper care of signs:
\[
\psi (x) = \dt e^{\textstyle\,i \sqrt \pi \gamma^5 [ \Sigma (x) + \eta (x)]} \dt \psi^{(0)} (x)\,,
\]
\be\label{pt}
\widetilde\psi (x) =  \dt e^{\textstyle\,  i \sqrt \pi \gamma^5 [ \widetilde\Sigma (x) + 
\widetilde\eta (x)]} \dt \widetilde\psi^{(0)} (x)\,,
\ee
\[
{\cal A}_\mu (x)\, =\, -\, \frac{\sqrt \pi}{e} \epsilon_{\mu \nu} \partial^\nu 
\Big ( \Sigma (x) + \eta (x) \Big )\,,
\]
\be\label{at}
\widetilde{\cal A}_\mu (x)\, =\, + \,\frac{\sqrt \pi}{e} \epsilon_{\mu \nu} \partial^\nu 
\Big ( \widetilde\Sigma (x) + \widetilde\eta (x) \Big )\,.
\ee

\no The fields $\Sigma$ and $\widetilde\Sigma$ are free pseudoscalar fields of mass $ m ={e}/{\sqrt\pi}$, and $\eta$, $\widetilde\eta$ are free massless fields quantized with indefinite metric \cite{LS}. The free Fermi fields are given in terms of the Wick-ordered exponentials of the free massless scalar fields by \cite{ABRII} 
\footnote{We have suppressed the Klein factors needed to ensure the correct anticommutation relations \cite{ABR}.},

\be\label{psif}
\psi^{(0)} (x) = \Big ( \frac{\mu}{2 \pi} \Big )^{\frac{1}{2}} \dt e^{\textstyle\,i \sqrt \pi [
\gamma^5 \phi (x) + \phi_D (x)]} \dt\,,
\ee
\be\label{psiftild}
\widetilde\psi^{(0)} (x) = \Big ( \frac{\mu}{2 \pi} \Big )^{\frac{1}{2}}
\dt e^{\textstyle\,i\, \sqrt \pi [
\gamma^5 \widetilde\phi (x) + \widetilde{\phi}_D (x)]} \dt\,,
\ee

\no where $\mu$ is an infrared regulator of the massless scalar fields $\phi$ and $\widetilde\phi$. The fields $\phi_D$ and $\widetilde\phi_D$ are the duals of $\phi$  and $\widetilde\phi$,

\[
\partial_\mu \phi_D (x)\,=\,\varepsilon_{\mu \nu } \partial ^\nu \phi (x)\,,
\quad
\partial_\mu \widetilde\phi_D (x)\,=\,\varepsilon_{\mu \nu } \partial ^\nu \widetilde\phi (x)\,.
\]

\no The two-point functions of the left- and right-moving components of the free massless scalar fields $\phi$ and $\widetilde\phi$ are  given by 
\footnote{For notational simplicity we have dropped the $l$ and $r$ subscripts in the decompositions
$\phi(x)=\phi_r(x^-)-\phi_l(x^+)$ and $\phi_D(x)=\phi_r(x^-)+\phi_l(x^+)$.}

\be
\langle 0  \vert \phi (x^\pm) \phi (0) \vert 0 \rangle\,=\,
-\,\frac{1}{4 \pi}\,\ln \big \{ i \mu ( x^\pm\,-\,i \epsilon ) \big \}\,,\label{phi2pf1}
\ee
\be\label{phi2pf2}
\langle \widetilde 0 , \vert \widetilde{\phi} (x^\pm) \widetilde{\phi} (0) \vert \widetilde 0 \rangle\, = 
-\,\frac{1}{4 \pi}\,\ln \big \{- i \mu ( x^\pm\,+\,i \epsilon ) \big \}\,.
\ee

The Wick-ordered exponential of a free boson field is defined 
in terms of the creation and annihilation field components as

\[
\dt e^{i\,\lambda \phi (x)}\dt \doteq e^{i\,\lambda \phi^{(-)} (x)}\,e^{i\,\lambda \phi^{(+)} (x)}\,.
\]

\no The bosonization of free massless fermions in two-dimensions is based on the fact that the correlation functions of Wick-ordered exponentials of the free massless scalar field satisfy Wightman positivity, provided we associate to the exponential a conserved charge $\lambda$ \cite{AAR,AARII,Swieca,W}. This superselection rule, contained in the exponentials of the  bosonized fermion theory, implements the fermion charge and pseudo charge conservation. As we shall see, the infrared regulator $\mu$, that appears in the bosonized expressions (\ref{psif}) and (\ref{psiftild}) plays an important role in order to ensure the fermion charge and chirality selection rules. The limit $\mu \rightarrow 0$ ensures that the only non-zero Wightman functions are those for which the bare charge and chirality are conserved. Following the spirit of Ref. \cite{Swieca}, this will be explained in detail later both in the $T = 0$ and $T \neq 0$ case.

For the massless free fields $\eta (x)$ and $\widetilde\eta (x)$, quantized with indefinite metric, we have \cite{LS}: 

\be\label{eta2pf1}
\langle 0  \vert \eta (x^\pm) \eta (0) \vert 0 \rangle\,=\,
\frac{1}{4 \pi}\,\ln \big \{ i \bar\mu ( x^\pm\,-\,i \epsilon ) \big \}\,,
\ee
\be \label{eta2pf2}
\langle \widetilde 0 , \vert \widetilde{\eta} (x^\pm) \widetilde{\eta} (0) \vert \widetilde 0 \rangle\,=\,
\frac{1}{4 \pi}\,\ln \big \{ -i \bar\mu ( x^\pm\,+\,i \epsilon ) \big \}\,.
 \ee
 
\no Here $\mu$ and $\bar\mu$ are independent infrared cutoffs. The fermion charge and chirality selection rules are carried by the Wick-ordered exponentials of the fields $\phi$ and $\widetilde\phi$. The indefinite metric fields $\eta$ and $\widetilde\eta$ do not carry these selection rules and the cutoff $\bar\mu$ will be maintained at a fixed finite,
though small, value.

Let us remark that the sign associated with both the $\widetilde\Sigma$ and $\widetilde\eta$ fields in equation (\ref{pt}), and whence in
eq. (\ref{at}), is dictated by the bosonized form of the free Fermi field (for more details see Refs. \cite{ABR,ABRII}).  Indeed, the bosonized expression (\ref{psiftild}) for the field $\widetilde\psi^{(0)} (x)$ is not obtained by the tilde conjugation operation, defining $\widetilde{(c \phi)} = c^\ast \widetilde\phi$, but rather by the tilde conjugation of just the  field $\phi (x)$ in the exponent \cite{ABRII}. This bosonization prescription for $\psi^{(0)}$ proves necessary  for obtaining agreement with the real-time formalism of Umezawa et al \cite{U}  for the off-diagonal two-point function $\langle  \widetilde\psi^{(0)} \psi^{(0)} \rangle$ at finite temperature, as well as agreement with the revised thermofield approach for fermions due to Ojima \cite{O}. For a more detailed discussion of the bosonization of the free Fermi thermofield we refer the reader to the Refs. \cite{ABR,ABRII}.

The vector currents are computed using the standard gauge invariant 
point-splitting limit prescription \cite{LS}

\[
{\cal J}^\mu (x) = \vdt \overline\psi (x) \gamma^\mu
\psi (x) \vdt\, \doteq \lim_{\varepsilon \rightarrow 0} f (\varepsilon )
\Bigg \{\overline\psi (x + \varepsilon)\,
e^{\textstyle\,\,i e \int_x^{x + \varepsilon} {\cal A}_\mu (z)\,d z^\mu}\,\gamma^\mu
\psi (x) - V. E. V. \Bigg \}\,,
\]
\[
\widetilde{\cal J}^\mu (x) = \vdt \overline{\widetilde\psi} (x) \gamma^\mu
\widetilde\psi (x) \vdt\, \doteq \lim_{\varepsilon \rightarrow 0} \widetilde f (\varepsilon )
\Bigg \{\overline{\widetilde\psi} (x + \varepsilon )\,
e^{\textstyle\,\,-\,i e \int_x^{x + \varepsilon} \widetilde{\cal A}_\mu (z)\,d z^\mu}
\widetilde\psi (x) - {V.E.V.} \Bigg \}\,,
\]

\no where $f (\epsilon )$ is a suitably chosen renormalization constant, and the limit is taken along the space-like direction. Using the two-point functions of the free massless scalar fields (\ref{phi2pf1})-(\ref{eta2pf2}), and 
the short-distance two-point functions of the free massive scalar fields \cite{ABR} ($\gamma$ is the Euler constant)

\[
[ \Sigma^{(+)} (x )\,,\,\Sigma^{(-)} (0) ]\,\approx\,-\,\frac{1}{4\pi}\,\ln \big [ \big ( m e^\gamma \big )^2\big (-x^2+i  \epsilon x^0 \big ) \big ]\,,
\]
\[
[ \widetilde\Sigma^{(+)} (x )\,,\,\widetilde\Sigma^{(-)} (0) ]\,\approx\,-\,\frac{1}{4\pi}\,\ln \big [ \big (m e^\gamma \big )^2\big (-x^2-i  \epsilon x^0 \big ) \big ]\,,
\]

\no we obtain \footnote{Note that the currents (\ref{cur1}) and (\ref{cur2}) are separately conserved. As one readily checks, this is in accordance with the Hamilton equations of motion generated by the total Hamiltonian of the combined system $\widehat H = H - \widetilde H$,

\[
\partial_0 {\cal J}_0 (x) = \big [ {\cal J}_0 (x)\,,\,\widehat H \big ]\,=\,\partial_1 {\cal J}_1 (x)\,.
\]}
\be\label{cur1}
{\cal J}_\mu (x) = \,-\, \frac{1}{\sqrt \pi}\,\epsilon_{\mu \nu} 
\partial^\nu \Sigma ( x )  + \lc_\mu (x)\,,
\ee
\be\label{cur2}
\widetilde{\cal J}_\mu (x) =\,+\, \frac{1}{\sqrt \pi}\,\epsilon_{\mu \nu} 
\partial^\nu \widetilde\Sigma ( x )  
+ \widetilde{\lc}_\mu (x)\,,
\ee

\no where $\lc_\mu (x)$ and $\widetilde{\lc}_\mu (x)$
are longitudinal zero mass contributions to the currents 

\[
\lc_\mu (x  ) = \,- \,\frac{1}{\sqrt \pi}\,\epsilon_{\mu \nu} \partial^\nu \varphi (x )  =
\,- \,\frac{1}{\sqrt \pi}\,\partial_\mu \varphi_D (x ) \,,
\]
\[
\widetilde{\lc}_\mu (x ) = \, \frac{1}{\sqrt \pi}\,\epsilon_{\mu \nu} \partial^\nu \widetilde\varphi (x) =
\,\frac{1}{\sqrt \pi}\,\partial_\mu \widetilde\varphi_D (x)
\,,
\]

\no and  $\varphi$, $\widetilde\varphi$  are fields acting as potentials for the longitudinal currents, defined by 

\be\label{varphi1}
\varphi (x) = \eta (x ) + \phi (x ) \,,
\quad
\widetilde\varphi (x) = \widetilde\eta (x ) + \widetilde\phi (x ) \,.
\ee

\no Due to the opposite metric quantization of the fields $\eta$ and $ \widetilde\eta$ with respect to $\phi$ and $\widetilde\phi$, the longitudinal currents ${\lc}_\mu$ and $\widetilde{\lc}_\mu$, both   generate zero norm states on the vacuum (\ref{Fock-vacuum})

\[
\langle 0 , \widetilde 0 \vert \lc_\mu (x) \lc_\nu (y ) 
\vert \widetilde 0 , 0 \rangle = 0\,,\,\,\,\forall (x , y)\,,
\]
\[
\langle 0 , \widetilde 0 \vert \widetilde{\lc}_\mu (x) 
\widetilde{\lc}_\nu (y) 
\vert \widetilde 0 , 0 \rangle = 0\,,\,\,\,\forall (x , y)\,.
\]

\no Note that, since at $T =  0$ the fields $\phi$ ($\eta$) and $\widetilde\phi$ ($\widetilde\eta$) are independent

\[
\langle 0 , \widetilde 0 \vert \eta (x) \widetilde\eta (y) \vert \widetilde 0 , 0 \rangle\,\equiv\,0\,,
\quad
\langle 0 , \widetilde 0 \vert \phi (x) \widetilde\phi (y) \vert \widetilde 0 , 0 \rangle\,\equiv\,0\,,
\]

\no we also have 

\[
\langle 0 , \widetilde 0 \vert {\lc}_\mu (x) 
\widetilde{\lc}_\nu (y) 
\vert \widetilde 0 , 0 \rangle \equiv 0\,,\,\,\,\forall (x , y)\,,
\]

\no regardless of the sign of the metric quantization of the fields.

The free fermionic currents are given by

\[
j^\mu (x)\,=\,-\,\frac{1}{\sqrt \pi}\,\epsilon^{\mu\nu}\,\partial_\nu {{\phi}} (x)\,=\,\epsilon^{\mu \nu} j^5_\nu (x)\,,
\] 
\[
\widetilde j^\mu (x)\,=\,+\,\frac{1}{\sqrt \pi}\,\epsilon^{\mu\nu}\,\partial_\nu {\widetilde{{\phi}}} (x)\,=\,\epsilon^{\mu \nu} \widetilde j^5_\nu (x)\,.
\]

\no The free charge and pseudo charge operators are formally defined by

\[
{\cal Q}_f\,=\,\int j_0 (z) dz^1\,\,\,\,,\,\,\,\,{\cal Q}^5_f\,=\,\int j^5_0 (z) dz^1\,,
\]
\[
\widetilde{\cal Q}_f\,=\,\int \widetilde j_0 (z) dz^1\,\,\,\,\,,\,\,\,\,\widetilde{\cal Q}^5_f\,=\,\int \widetilde j^5_0 (z) dz^1\,,
\]

\no and using the equal-time commutation relations \cite{ABRII}

\[
[ {{\phi_D}} (x)\,,\,\partial_0 {{\phi_D}} (y) ]\,=\,i\,\delta (x^1\,-\,y^1)\,,
\]
\[
[ \widetilde\phi_D (x)\,,\,\partial_0 \widetilde\phi_D (y) ]\,=\,-\,i\,\delta (x^1\,-\,y^1)\,,
\]

\no  one finds that $\widetilde\psi^{(0)}$ is an identical copy of $\psi^{(0)}$ carrying the same charge and chirality quantum numbers

\[
[ {\cal Q}_f\,,\,\psi^{(0)} (x) ]\,=\,-\,\psi^{(0)} (x)\,,\,\,\,\,\,\,\,\,\,\,\,\,[ {\cal Q}^5_f\,,\,\psi^{(0)} (x) ]\,=\,-\,\gamma^5\,\psi^{(0)} (x)\,,
\]
\[
[ \widetilde{\cal Q}_f\,,\,\widetilde\psi^{(0)} (x) ]\,=\,-\,\widetilde\psi^{(0)} (x)\,,\,\,\,\,\,\,\,\,\,\,\,\,[ \widetilde{\cal Q}^5_f\,,\,\widetilde\psi^{(0)} (x) ]\,=\,-\,\gamma^5\,\widetilde\psi^{(0)} (x)\,.
\]

\no In thermofield dynamics, the above algebraic relations will be retained at finite temperature.

As in the standard model \cite{LS}, physical  (gauge invariant) states $\vert\Phi \rangle $ , $ \vert \widetilde\Phi \rangle $, are defined by the subsidiary conditions 

\[
\langle \Phi^\prime  \vert \lc_\mu (x) \vert  \Phi \rangle\,=\,0\,, 
\quad
\langle  \widetilde\Phi^\prime \vert \widetilde{\lc}_\mu (x) \vert \widetilde\Phi \rangle\,=\,0\,.
\]

\no On the physical states  Maxwell's equations are satisfied in the weak sense:

\[
\langle \Phi^\prime \vert \big ( \partial_\mu {\cal F}^{\mu \nu } (x)\,+\,e\,{\cal J}^\nu (x) \big ) \vert  \Phi \rangle \,=\,0\,,
\]
\[
\langle  \widetilde\Phi^\prime \vert \big ( \partial_\mu \widetilde{\cal F}^{\mu \nu } (x)\,+\,e\,\widetilde{\cal J}^\nu (x) \big ) \vert \widetilde\Phi \rangle \,=\,0\,.
\]

\no The physical Hilbert subspace ${\hat{\cal H}}_{phys}$ is obtained by applying on the Fock vacuum 
$\vert \widetilde 0 , 0 \rangle = |0>\otimes|0>$  Wightman polynomials of operators $\{{\cal O}\}$ that are {\it strictly gauge invariant} \cite{Sym}, i. e.,

\be\label{opphys}
\big [ \{{\cal O}\}\,,\,\lc_\mu (x) \big ]\,=\,0\,\,\,\,,\,\,\,\,
\big [ \{{\cal O}\}\,,\,\widetilde{\lc}_\mu (x) \big ]\,=\,0\,.
\ee

\no At $T = 0$ the set of operators $\{\Phi\}$ and $\{\widetilde\Phi\}$ are independent, and a general gauge invariant  state is given by 

\[
\vert \widetilde\Phi , \Psi \rangle\,=\,\vert \widetilde\Phi \rangle \otimes \,\vert \Psi \rangle\,=\,\widetilde\Phi \vert \widetilde 0 \rangle \otimes \Psi \vert 0 \rangle\,. 
\]

\no In the zero temperature case one has the factorization of the physical Hilbert space 

\[
{\hat{\cal H}}_{phys}\,=\,{\cal H}_{phys} \otimes \widetilde{\cal H}_{phys}\,.
\]

\no As we shall see this will be no longer true in the finite temperature case.

\subsection{Gauge Invariant Field Algebra}

The physical states are  obtained by applying gauge invariant operators on the Fock vacuum state. Such operators are $\{{\cal F}^{\mu\nu}, \widetilde{\cal F}^{\mu \nu}, {\cal J}^\mu, \widetilde{\cal J}^\mu \}$, as well as the ``dipole" operators formally defined by \cite{LS}

\[
D (x , y)\,\sim\,\psi (x)\,e^{\,i\,e\,\int_x^y {\cal A}_\mu (z) d z^\mu}\,\psi^\dagger (y)\,,
\]
\[
\widetilde D (x , y)\,\sim\,\widetilde \psi (x)\,e^{\,-\,i\,e\,\int_x^y \widetilde{\cal A}_\mu (z) d z^\mu}\,\widetilde\psi^\dagger (y)\,.
\]

\no Using the gauge freedom of the theory, one may consider  the set of gauge invariant field operators $\{ \Psi, \widetilde\Psi, \mathbf{A}_\mu, \widetilde\mathbf{A}_\mu \}$, related to the set of fields $\{\psi, \widetilde\psi, {\cal A}_\mu, \widetilde{\cal A}_\mu \}$, by the operator-valued gauge transformations \cite{LS,Sym} (the gauge $\sqrt\pi$ of Ref. \cite{LS}) 

\be\label{ph}
\Psi (x)\,=\,\dt \psi (x)\,e^{\displaystyle\,i\,\sqrt\pi \,\eta_{_D} (x)}\dt\,=\,\big ( \frac{\overline{\mu}}{2 \pi} \big )^{\frac{1}{2}}\,\dt e^{\displaystyle\,i\,\sqrt\pi\,\gamma^5 \Sigma (x)}\dt \,\sigma (x)\,,
\ee
\be\label{tph}
\widetilde\Psi (x)\,=\,\dt \widetilde\psi (x)\,e^{\displaystyle\,-\,i\,\sqrt\pi\,\widetilde\eta_{_D} (x)} \dt\,=\,\big ( \frac{\overline{\mu}}{2 \pi} \big )^{\frac{1}{2}}\,\dt e^{\displaystyle\,i\,\sqrt\pi\,\gamma^5 \widetilde\Sigma (x)}\dt \,\widetilde\sigma (x)\,,
\ee
\[
\mathbf{A}_\mu (x)\,=\,{\cal A}_\mu (x) +\,\frac{1}{e} \partial_\mu \eta_{_D} (x)\,=\,-\,\frac{\sqrt\pi}{e}\,\epsilon_{\mu \nu} \partial^\nu \Sigma (x)\,,
\]
\[
\widetilde{\mathbf{A}}_\mu (x)\,=\,\widetilde{\cal A}_\mu (x)-\,\frac{1}{e}\,\partial_\mu \widetilde\eta_{_D} (x)\,=\,
+\,\frac{\sqrt\pi}{e}\,\epsilon_{\mu \nu} \partial^\nu \widetilde\Sigma (x)\,,
\]

\no where we have defined the dimensionless operators \cite{LS}

\be\label{sig}
\sigma_\alpha (x)\,=\,e^{\displaystyle\,2\,i\,\sqrt\pi\,\varphi (x^\pm) }\,
=\,\big(\frac{\mu}{\bar\mu}\big)^\frac12\,\dt\,
e^{\displaystyle\,i\,\sqrt\pi\,\big ( \gamma^5_{\alpha\alpha}  \varphi (x) \,+ \, \varphi_{_D} (x)\big ) } \dt\,=\,\big(\frac{\mu}{\bar\mu}\big)^\frac12\,\dt\,e^{\displaystyle\,2\,i\,\sqrt\pi\,\varphi (x^\pm) }\dt\,
\ee
\be\label{sigt}
\widetilde\sigma_\alpha (x)\,=\,e^{\displaystyle\,2\,i\,\sqrt\pi\,\widetilde\varphi (x^\pm)} \,=\,(\frac{\mu}{\bar\mu})^\frac12\,
\dt 
e^{\displaystyle\,i\,\sqrt\pi\,\big ( \gamma^5_{\alpha\alpha}  \widetilde\varphi (x)\,+\, \widetilde{\varphi}_{_D} (x) \big )}\dt\,=\,(\frac{\mu}{\bar\mu})^\frac12\,\dt\,
e^{\displaystyle\,2\,i\,\sqrt\pi\,\widetilde\varphi (x^\pm)\,}\,\dt
\ee

\no and $x^\pm$ refers to the distinct chirality labels of the $\sigma_{\alpha}$ fields, $\sigma_1=\sigma(x^+)$ and $\sigma_2=\sigma(x^-)$. Note that for $T=0$ the tilde-operators just play the role of ``spectators''. Analogous to the standard Schwinger model, the operators $\sigma$ and $\tilde\sigma$ are constant operators that commute among themselves and
with all gauge invariant operators \cite{LS}. The only reason for $\sigma$ and $\widetilde\sigma$ not being the identity  operator on the Hilbert space is that they carry a selection rule corresponding to the
charge and the chirality associated to each free fermion field $\psi^{(0)}$ and $\widetilde\psi^{(0)}$,

\be\label{a1}
\big [ {\cal Q}_f\,,\,\sigma \big ]\,=\,-\,\sigma\,,\,\,\,\,\,\,\,\,\,\,\, \big [ {\cal Q}^5_f\,,\,\sigma \big ]\,=\,-\,\gamma^5\,\sigma\,,
\ee
\be\label{a4}
\big [ \widetilde{\cal Q}_f\,,\,\widetilde\sigma \big ]\,=\,-\,\widetilde\sigma\,,\,\,\,\,\,\,\,\,\,\,\,\,\big [ \widetilde{\cal Q}^5_f\,,\,\widetilde\sigma \big ]\,=\,-\,\gamma^5\,\widetilde\sigma\,.
\ee

\subsection{Superselection Rule and Vacuum Structure}

In order to display in the zero temperature case the role played by the superselection rules carried by the Wick-ordered exponentials of the free Bose fields ($\phi$,$\widetilde\phi$) in (\ref{psif})-(\ref{psiftild}), let us compute the general Wightman functions of the operators $\sigma$ and $\widetilde\sigma$. To this end we shall make use of the two-point functions (\ref{phi2pf1})-(\ref{eta2pf2}), implying for the fields $\varphi$ and $\widetilde\varphi$ in (\ref{varphi1}) the constant correlation functions

\be\label{vphivphi}
\langle 0 , \widetilde 0 \vert \varphi (x) \varphi (y) \vert \widetilde 0 , 0 \rangle\,=\,
\langle 0 , \widetilde 0 \vert \widetilde\varphi (x) \widetilde\varphi (y) \vert \widetilde 0 , 0 \rangle\,=\,
-\,\frac{1}{4 \pi}\,\ln \frac{\mu}{\overline\mu}\,,
\ee
\be\label{vphitildevphi}
\langle 0 , \widetilde 0 \vert \widetilde\varphi (x) \varphi (y) \vert \widetilde 0 , 0 \rangle\,=\,0\,.
\ee

\no Hence the operators $\sigma$ and $\widetilde\sigma$,  defined by (\ref{sig}) and (\ref{sigt}), generate constant Wightman functions. Since at $T = 0$ the fields $\varphi$ and $\tilde\varphi$ are independent, as expressed by (\ref{vphitildevphi}), there are no correlations between the fields $\sigma$'s and $\tilde\sigma$'s,  and a general Wightman functions factorizes into a direct product. Using (\ref{vphivphi}), as  a consequence of the bare charge and chirality selection rules carried by the Wick-ordered exponentials of the free massless fields $\phi (x)$ and $\widetilde\phi (x)$, the  Wightman functions are given by \cite{LS}

$$
\langle 0 \vert \sigma_1 (x_1) \cdots \sigma_1 (x_{m_1}) \sigma_2 (y_{1}) \cdots \sigma_2 (y_{m_2}) \sigma_1^\dagger (x^\prime_1) \cdots \sigma_1^\dagger (x^\prime_{m^\prime_1}) \sigma_2^\dagger (y^\prime_{1})\cdots \sigma_2^\dagger (y^\prime_{m^\prime_2}) \vert 0 \rangle 
$$
\be\label{wfsigma}
=\,\lim_{\mu \rightarrow 0}\,\Bigg (\frac{\mu}{\overline\mu} \Bigg )^{\displaystyle\, \frac{1}{2}\big [ ( m_1\,-\,m^\prime_1 )^2\,+\,( m_2\,-\,m^\prime_2 )^2\,\big ]}\,=\,
\delta_{m_1 -\,m^\prime_1\,,\,0}\cdot\delta_{m_2\,-\,m^\prime_2\,,\,0}\ \ \,
\ee
$$
\langle \widetilde 0 \vert \widetilde\sigma_1 (\tilde x_1) \cdots \widetilde\sigma_1 (\tilde x_{\tilde m_1}) \widetilde\sigma_2 (\tilde y_{1}) \cdots \widetilde\sigma_2 (\tilde y_{\tilde m_2}) \widetilde\sigma_1^\dagger (\tilde x^\prime_1) \cdots \widetilde\sigma_1^\dagger (\tilde x^\prime_{\tilde m^\prime_1}) \widetilde\sigma_2^\dagger (\tilde y^\prime_{1})\cdots \widetilde\sigma_2^\dagger (\tilde y^\prime_{\tilde m^\prime_2}) \vert \widetilde 0 \rangle
$$
\be\label{wftildesigma}
=\,\lim_{\mu \rightarrow 0}\,\Bigg (\frac{\mu}{\overline\mu} \Bigg )^{\displaystyle\, \frac{1}{2}\big [ \,( \tilde m_1\,-\,\tilde m^\prime_1  )^2\,+\,( \tilde m_2\,-\,\tilde m^\prime_2 )^2 \big ]}\,=\,\delta_{\tilde m_1\,-\,\tilde m^\prime_1\,,\,0}\cdot\delta_{\tilde m_2\,-\,\tilde m^\prime_2\,,\,0}\,.
\ee

\no For each spinor field component $\sigma_\alpha$, the total charge ($q_\alpha$) and chirality ($q^5_\alpha$) quantum numbers carried by the Wightman functions above are given by

\be\label{sel1}
q_\alpha\,=\,m^\prime_\alpha - m_\alpha\,\,\,\,,\,\,\,\,
q^5_\alpha\,=\,\gamma^5_{\alpha\alpha} \,(m^\prime_\alpha - m_\alpha)\,,
\ee
\be\label{sel2}
\tilde q_\alpha\,=\,\tilde m^\prime_\alpha - \tilde m_\alpha\,\,\,\,,\,\,\,\,\tilde q^5_\alpha\,=\,\gamma^5_{\alpha\alpha}\,(\tilde m^\prime_\alpha - \tilde m_\alpha )\,.
\ee

\no In the zero temperature case the selection rules require two independent conservation laws, implying that the only non-zero Wightman functions, which also are cutoff independent,  are those  with $q_\alpha  = 0$ and $\tilde q_\alpha = 0 $. The cluster property is violated independently for each of the Hilbert spaces ${\cal H}$ and $\widetilde{\cal H}$. The Hilbert spaces ${\cal H}$ and $\widetilde{\cal H}$  generated cyclically from the Fock vacua $\vert 0 \rangle$ and $ \vert \widetilde 0 \rangle$ will contain an infinite number of vacuum states \cite{LS}. As in the case of the usual Schwinger model \cite{LS}, the $\sigma$'s and $\widetilde\sigma$'s form an Abelian algebra of constant unitary operators

\[
\sigma_\alpha (x) = \sigma_\alpha (0) \doteq \sigma_\alpha\,,
\quad
\widetilde\sigma_\alpha (x) = \widetilde\sigma_\alpha (0) \doteq \widetilde\sigma_\alpha\,,
\]
\[
\sigma_\alpha \sigma_\alpha^\dagger = 1 = \widetilde\sigma_\alpha \widetilde\sigma_\alpha^\dagger\,.
\]

\no Since the Hamiltonians $H$ and $\widetilde H$  commute with $\sigma$ and $\widetilde\sigma$,

\be\label{a5}
\big [ H\,,\,\sigma \big ]\,=\,\big [ H\,,\,\widetilde\sigma \big ]\,=\,\big [\widetilde H\,,\,\sigma \big ]\,=\,\big [ \widetilde H\,,\,\widetilde\sigma \big ]\,=\,0\,,
\ee

\no the ground state is degenerate. The infinite number of mutually orthogonal vacua being generated by repeated application of both, $\sigma_\alpha$ and $\widetilde\sigma_\alpha$ to the Fock-vacuum $
\vert \widetilde 0 , 0 \rangle\,=\,\vert \widetilde 0 \rangle \otimes \vert 0 \rangle\,\equiv\,\vert \widetilde 0_1 , \widetilde 0_2 \rangle\,\otimes \,\vert 0_1 , 0_2 \rangle\,,
$ ($1,2,$ stand for left- and right-chiral components) are 

\be\label{vacuum-states}
\vert \widetilde n_1 , \widetilde n_2 ; n_1 , n_2 \rangle \,=\,\widetilde \sigma_1^{\tilde n_1} \widetilde\sigma_2^{\tilde n_2}\,\vert \widetilde 0 , \widetilde 0 \,\rangle\,\otimes\,
\sigma_1^{n_1} \sigma_2^{n_2}\,\vert 0 , 0 \rangle\,=\,\vert \widetilde n_1 , \widetilde n_2 \rangle \otimes \vert n_2 , n_2 \rangle\,,
\ee
and satisfy
\[
H\,\vert \widetilde n_1 , \widetilde n_2 ; n_1 , n_2 \rangle \,=\,\widetilde H\,\vert \widetilde n_1 , \widetilde n_2 ; n_1 , n_2 \rangle \,=\,0\,.
\]

Equations (\ref{wfsigma}) and (\ref{wftildesigma}) demonstrate the orthogonality of the states (\ref{vacuum-states}) in accordance with the free fermion charge and chirality selection rules:

\be\label{sr0}
\langle  n^\prime_1 , n^\prime_2 ; \widetilde n^\prime_1 , \widetilde n^\prime_2 \vert \widetilde n_1 , \widetilde n_2 ; n_1 , n_2 \rangle\,=\,
\langle  n^\prime_1 , n^\prime_2 \vert  n_1 , n_2 \rangle\,\times\,
\langle   \widetilde n^\prime_1 , \widetilde n^\prime_2 \vert \widetilde n_1 , \widetilde n_2 \rangle\,=\,\delta_{\tilde n^\prime_1 - \tilde n_1 , 0} \cdot \delta_{\tilde n^\prime_2 - \tilde n_2 , 0} \times \delta_{ n^\prime_1 - n_1 , 0} \cdot \delta_{ n^\prime_2 -  n_2 , 0}\,.
\ee

\no Since in the zero temperature case the Hilbert space factorizes, as a consequence of the fermion charge and chirality selection rules, one has

\[
\langle 0 , \widetilde 0 \vert \sigma \tilde\sigma \vert \widetilde 0 , 0 \rangle\,=\,
\langle 0 \vert \sigma \vert 0 \rangle \langle \widetilde 0 \vert \tilde\sigma \vert \widetilde 0 \rangle\,=\,0\,.
\]

As in the standard treatment of the $T=0$ model, the observables commute with the fields $\sigma$ and $\widetilde\sigma$ for all space-time separations. Following Ref.\cite{LS},  we can introduce a set of irreducible vacuum states with respect to which the observables are diagonal, as follows

\be\label{thetavacuum}
\vert \theta_1 , \theta_2 ; \widetilde\theta_1 , \widetilde\theta_2 \rangle\,=\, \frac{1}{4 \pi^2}\,\sum_{n_1,n_2 = - \infty}^{+ \infty}\,e^{\,-\,i\,n_1 \theta_1}\,e^{\,-\,i\,n_2 \theta_2}\,\sum_{\tilde n_1,\tilde n_2 = - \infty}^{+ \infty}\,e^{\,\,i\,\tilde n_1 \widetilde\theta_1}\,e^{\,\,i\,\tilde n_2 \widetilde\theta_2}\,\vert n_1 , n_2 ; \tilde n_1 , \tilde n_2 \rangle\,,
\ee

\no such that the $\sigma$ and $\widetilde\sigma$ fields turn out to be ``spurionized" \cite{AAR}

\be\label{spurionized}
\sigma_\alpha\,\vert \theta_1 , \theta_2 ; \widetilde\theta_1 , \widetilde\theta_2 \rangle\,=\,e^{\,i\,\theta_\alpha}\,\vert \theta_1 , \theta_2 ; \widetilde\theta_1 , \widetilde\theta_2 \rangle\,,
\ee
\[
\widetilde\sigma_\alpha\,\vert \theta_1 , \theta_2 ; \widetilde\theta_1 , \widetilde\theta_2 \rangle\,=\,e^{\,-\,i\,\widetilde\theta_\alpha}\,\vert \theta_1 , \theta_2 ; \widetilde\theta_1 , \widetilde\theta_2 \rangle\,.
\]

\no In order to simplify the notation we shall write $\theta = \{\theta_1 , \theta_2\}$ , $\widetilde\theta = \{\widetilde\theta_1 , \widetilde\theta_2\}$, $\vert n_1 , n_2 \rangle = \vert n \rangle$, etc. As a consequence of the two independent selection rules (\ref{sel1})-(\ref{sel2}), the decoupling between distinct $\theta$-sectors is complete ($\delta^{(2)} (\theta) \equiv \delta (\theta_1)\, \delta (\theta_2)$)

\[
\langle \theta^\prime,  \widetilde\theta^\prime \vert \theta , \widetilde\theta \rangle\,=\,\frac{1}{(2 \pi)^4} \sum_{\tilde n, \tilde n^\prime} \sum_{n, n^\prime}\,e^{- i ( n \theta - n^\prime \theta^\prime )}\,e^{\,i\,(\tilde n \widetilde\theta - \tilde n^\prime \widetilde\theta^\prime ) }\,\delta_{n , n^\prime} \delta_{\tilde n , \tilde n^\prime}\,=\,\delta^{(2)} (\theta^\prime -\theta)\,
\delta^{(2)} (\widetilde\theta^\prime - \widetilde\theta)\,,
\]

\no such that we have two independent theta vacua sectors,

\[
\vert \theta ; \widetilde\theta \rangle\,=\,\vert \theta  \rangle \otimes \vert  \widetilde\theta \rangle\,.
\]

\no The irreducible representations of the field algebras $\mathbf{A}$ and $\widetilde\mathbf{A}$ are obtained by defining averaged vacuum expectation values of arbitrary Wightman polynomials ${\cal P} = {\cal O}  \otimes \widetilde{\cal O}$ of gauge invariant operators \cite{LS} by,

\be\label{vevtheta}
\overline{\langle 0 , \widetilde 0 \vert\,{\cal P}\,\vert \widetilde 0 , 0 \rangle}\,\doteq\,\int_0^{2\pi} d^2 \theta^\prime  \, \int_0^{2\pi} d^2 \widetilde\theta^\prime
\,\langle \theta^\prime  ;\widetilde\theta^\prime \vert \,\widetilde{\cal P}\,\vert\,\widetilde\theta ; \theta \rangle\,,
\ee
\[
=\,\int_0^{2\pi} d ^2 \theta^\prime \langle \theta^\prime \vert {\cal O} \vert \theta \rangle \times 
\int_0^{2\pi} d ^2 \widetilde\theta^\prime\,\langle \widetilde\theta^\prime \vert \,\widetilde{\cal O}\,\vert\,\widetilde\theta\rangle\,
\]
\[
=\,\overline{\langle 0 \vert\,{\cal O}\,\vert  0 \rangle}\,\otimes\,\overline{\langle \widetilde 0 \vert\,\widetilde{\cal O}\,\vert \widetilde 0  \rangle}
\]

\no In this $\theta$-vacuum representation one has

\[
\overline{\langle 0 , \widetilde 0 \vert\,\sigma_\alpha\,\widetilde\sigma_\gamma \,\vert \widetilde 0 , 0 \rangle}\,=\,\overline{\langle 0 \vert\,\sigma_\alpha\,\vert  0 \rangle}\,\otimes\,\overline{\langle \widetilde 0 \vert\,\widetilde\sigma_\gamma\,\vert \widetilde 0  \rangle}\,=\,e^{\,i\,(\theta_\alpha\,-\,\widetilde\theta_\gamma )}\,,
\]

\no displaying the spontaneous symmetry breaking in the physical Hilbert space ${\cal H}_{_{phys}} \otimes \widetilde{\cal H}_{_{phys}}$.

In the zero temperature case, since there are no correlations between the original system under consideration and the  tilde system, there is no a priori reason to require that $\theta = \widetilde\theta$. However, if we require the symmetry under the tilde conjugation rule one should redefine the theta vacuum expectation value (\ref{vevtheta}) as

\[
\overline{\overline{\langle 0 , \widetilde 0 \vert\,{\cal P}\,\vert \widetilde 0 , 0 \rangle}}\,\doteq\,\int_0^{2\pi} d^2 \theta^\prime  \, \int_0^{2\pi} d^2 \widetilde\theta^\prime\,\Big [ \int_0^{2\pi} d ^2 \widetilde\theta\,\,\delta^{(2)} (\widetilde\theta\,-\,\theta ) \Big ]\,
\,\langle \theta^\prime  ;\widetilde\theta^\prime \vert \,\widetilde{\cal P}\,\vert\,\widetilde\theta ; \theta \rangle\,,
\]

\no such that

\[
\overline{\overline{\langle 0 , \widetilde 0 \vert\,\sigma_\alpha\,\vert \widetilde 0 , 0 \rangle}}\,=\,e^{\,i\,\theta_\alpha}\,\,\,\,\,,\,\,\,\,\,
\overline{\overline{\langle 0 , \widetilde 0 \vert\,\widetilde\sigma_\alpha\,\vert \widetilde 0 , 0 \rangle}}\,=\,e^{\,-\,i\,\theta_\alpha}\,=\,\widetilde{\big (e^{i\, \theta_\alpha}\big )}\,.
\]

\no As we shall see, at finite temperature the tilde conjugation rule is ensured by the thermal selection rule.

\section{$QED_2$ at Finite Temperature}

\setcounter{equation}{0}

The above considerations have set the stage for introducing the temperature dependence.
To this end we depart from the operator solution for the combined doubled system at $T = 0$, as given in terms of the independent free Bose fields $\Sigma$, $\eta$, $\phi$, and their corresponding tilde fields. For $T \neq 0$, the statistical-mechanical ensemble averages are then given by
the expectation values of observables in a temperature dependent vacuum,  $\vert 0 (\beta ) \rangle$ \cite{O,U,TFD}, obtained via a Bogoliubov transformation from the $T=0$ Fock vacuum
$\vert 0,\tilde 0 \rangle$. These thermal averages factorize into  products of 
independent thermal averages corresponding
to the distinct free Bose field sectors. Let us thus introduce the creation and annihilation operators, acting on the Fock vacuum, $(a^\dagger,a)$, $(\widetilde a^\dagger, \widetilde a)$, $(b^\dagger, b)$, $(\widetilde b^\dagger, \widetilde b)$, $(c^\dagger, c)$ and $(\widetilde c^\dagger, \widetilde c)$ for the fields $\Sigma$, $\widetilde\Sigma$, $\phi$, $\widetilde\phi$, $\eta$ and $\widetilde\eta$, respectively.  The 
thermal vacuum is then defined in terms of  independent 
Bogoliubov transformations as follows,

\[
\vert 0 (\beta ) \rangle = {\cal U} ( \beta ) \vert \widetilde 0 , 0 \rangle\,=\, {\cal U}_\Sigma [\vartheta_\Sigma (\beta )] {\cal U}_\eta [\vartheta (\beta )]\, {\cal U}_\phi [\vartheta (\beta )]\, \vert \widetilde 0 , 0 \rangle\,.
\]

\no where the unitary operators above 
are given in terms of the 
creation and annihilation operators of the fields $(\Sigma,\widetilde\Sigma)$, 
$(\phi, \widetilde\phi)$ and $(\eta, \widetilde\eta)$,

\[
{\cal U}_\Sigma [\vartheta_\Sigma (\beta )]  = e^{\textstyle\,\,\int_{- \infty}^{+ \infty} dp^1\,
\Big (a^\dagger (p^1) {\widetilde a}^\dagger (p^1) - a (p^1) \widetilde a (p^1) 
\Big ) \vartheta_\Sigma ( p^0 ; \beta )}\,,
\]

\no with $p^0 =\,\sqrt{p_1^2\,+\,m^2}$, and 

\[
{\cal U}_\phi [\vartheta (\beta )]  = e^{\textstyle\,\,\int_{- \infty}^{+ \infty} dp^1\,
\Big (b^\dagger (p^1) {\widetilde b}^\dagger (p^1) - b(p^1) \widetilde b(p^1) 
\Big ) \vartheta (\vert p^1\vert ; \beta )}\,,
\]
\be\label{u3}
{\cal U}_\eta [\vartheta (\beta )]  = e^{\textstyle\,\,-\,\int_{- \infty}^{+ \infty} dp^1\,
\Big (c^\dagger (p^1) {\widetilde c}^\dagger (p^1) - c(p^1) \widetilde c(p^1) 
\Big ) \vartheta (\vert p^1\vert ; \beta )}\,.
\ee

\no The minus sign in the exponential in (\ref{u3}) is due to indefinite metric
quantization for the fields $\eta$ and $\widetilde\eta$ \cite{O},

\[
[ c (p^1) , c^\dagger (k^1) ]\, =\,-\, \delta (p^1 - k^1)\,.
\]

\no The Bogoliubov parameters are implicitly defined by \cite{Das,U,TFD}

\[
\sinh \vartheta_\Sigma ( p^0  \beta ) = 
\frac{\displaystyle e^{\displaystyle -\, \frac{\beta}{2}  p^0 }}
{\displaystyle\sqrt{1 - e^{\displaystyle -\, \beta  p^0 }}}\,=\,e^{ \displaystyle - \,\frac{\beta}{2}  p^0 }\,
\cosh \vartheta_\Sigma ( \beta p^0 )\,,
\]
\[
\sinh \vartheta (\vert p^1 \vert  \beta ) = 
\frac{e^{\displaystyle -\, \frac{\beta}{2} \vert p^1 \vert}}
{\sqrt{1 - e^{ \displaystyle -\, \beta \vert p^1 \vert}}}\,=\,
e^{\displaystyle -\, \frac{\beta}{2} \vert p^1 \vert}\,\cosh \vartheta (\beta \vert p^1 \vert ) \,,
\]

\no and the corresponding Bose-Einstein statistical weights are given by,

\[
N_\Sigma( p^0  ; \beta ) = \sinh^2 \vartheta_\Sigma ( \beta p^0 ) =
\frac{1}{\displaystyle e^{\displaystyle\beta  p^0 } - 1}\,,
\]
\[
N (\vert p^1 \vert ; \beta ) = \sinh^2 \vartheta (\beta\vert p^1 \vert ) =
\frac{1}{e^{\displaystyle\beta \vert p^1 \vert} - 1}\,.
\]

\no For any free bosonic field the corresponding  Bogoliubov transformed annihilation operators  ($e (p^1) = \{a,b,c\}(p^1)$) are given by \cite{Das,TFD,U}

\[
e (p^1 ; \beta )\,=\,{\cal U} [-\,\vartheta ( \beta )]\,e (p^1)\,{\cal U} [ \vartheta (\beta ) ]\,=\,e (p^1) \cosh \vartheta (p^0 ; \beta )\,-\,\tilde e^\dagger (p^1) \sinh \vartheta (p^0 ; \beta  )\,,
\]
\[
\tilde e (p^1 ; \beta )\,=\,{\cal U} [-\,\vartheta ( \beta )]\,\tilde e (p^1)\,{\cal U} [ \vartheta (\beta ) ]\,=\,\tilde e (p^1) \cosh \vartheta (p^0 ; \beta )\,-\,e^\dagger (p^1) \sinh \vartheta (p^0 ; \beta  )\,.
\]

\subsection{Gauge Invariant  Thermal Averages}

As in the zero temperature case, in the extension of thermofield dynamics formalism to gauge theories \cite{O}, the physical content of the model at finite temperature lies in the gauge invariant subspace of states. The statistical-mechanical ensemble averages of physical relevance are those expressed as expectation values with respect to physical thermal states.

To begin with, let $\{\Phi\}$ be the set of  gauge invariant operators at $T = 0$, as defined by (\ref{opphys}). At finite temperature a physical state is defined in terms of the temperature-dependent vacuum state $\vert 0 (\beta ) \rangle$ as

\[
\vert \Phi (\beta ) \rangle \doteq \Phi \vert 0 (\beta ) \rangle = {\cal U} (\beta) \Phi (\beta ) \vert \widetilde 0 , 0 \rangle\,.
\]
where
\[
\Phi(\beta) = U^{-1}(\beta)\Phi U(\beta)
\]

\no The physical (gauge invariant) Hilbert space $h_{_{phys}} (\beta ) = \{\vert \Phi (\beta ) \rangle \}$ at finite temperature is now defined by the  subsidiary conditions

\[
\langle \Phi^\prime (\beta ) \vert \lc_\mu (x) \vert \Phi (\beta ) \rangle = \langle 0 , \widetilde 0 \vert \Phi^\prime (\beta ) \lc_\mu (x ; \beta ) \Phi (\beta ) \vert \widetilde 0 , 0 \rangle\,=\,0\,,
\]
\[
\langle \Phi^\prime (\beta ) \vert \widetilde{\lc}_\mu (x) \vert \Phi (\beta ) \rangle = \langle 0 , \widetilde 0 \vert \Phi^\prime (\beta ) \widetilde{\lc}_\mu (x ; \beta ) \Phi (\beta ) \vert \widetilde 0 , 0 \rangle\,=\,0\,,
\]

\no where $\lc_\mu (x ; \beta )$ and $\widetilde{\lc}_\mu (x ; \beta)$ 
are the zero norm thermal longitudinal currents

\[
\lc_\mu (x ; \beta ) = \,- \,\frac{1}{\sqrt \pi}\,\epsilon_{\mu \nu} \partial^\nu 
 \varphi (x ; \beta ) =
\,- \,\frac{1}{\sqrt \pi}\,\partial_\mu \varphi_D(x;\beta) ,
\]
\[
\widetilde{\lc} (x ; \beta ) = \, \frac{1}{\sqrt \pi}\,\epsilon_{\mu \nu} \partial^\nu 
\widetilde\varphi (x ; \beta )   =
\, \frac{1}{\sqrt \pi}\,\partial_\mu \widetilde\varphi_D(x;\beta) \,,
\]

\no with

\be\label{vphibeta}
\varphi(x;\beta) = \phi(x;\beta) + \eta(x;\beta)\,,
\ee

\no and similarly for the other combinations, as well as the tilde field $\widetilde\varphi (x ; \beta) $. The thermal  vector currents are

\[
{\cal J}_\mu (x ; \beta ) = \,-\, \frac{1}{\sqrt \pi}\,\epsilon_{\mu \nu} 
\partial^\nu \Sigma ( x ; \beta )  + \lc_\mu (x ; \beta )\,,
\]
\[
\widetilde{\cal J}_\mu (x ; \beta ) =\, +\, \frac{1}{\sqrt \pi}\,\epsilon_{\mu \nu} 
\partial^\nu \widetilde\Sigma ( x ; \beta )  
+ \widetilde{\lc}_\mu (x ; \beta )\,,
\]

\no and the  Maxwell's equations are satisfied in the weak form

\[
\langle \Phi^\prime (\beta ) \vert \big ( \partial_\mu {\cal F}^{\mu \nu } (x)\,+\,e\,{\cal J}^\nu (x) \big ) \vert \Phi (\beta ) \rangle \,=\,0\,,
\]
\[
\langle \Phi^\prime (\beta ) \vert \big ( \partial_\mu \widetilde{\cal F}^{\mu \nu } (x)\,+\,e\,\widetilde{\cal J}^\nu (x) \big ) \vert \Phi (\beta ) \rangle \,=\,0\,,
\]

\no with 

\[
\left(\Box +\frac {e^2}{\pi}
\right)\Sigma (x;\beta)\,=\,\left(\Box +\frac {e^2}{\pi}
\right)\widetilde\Sigma (x;\beta)\,=\,0\,.
\]

\no The two-point functions of the  free massive thermofields  are discussed in detail in the Appendix. The field combination (\ref{vphibeta}) generates constant correlation functions. The two-point functions of the free massless thermofields are given by  \cite{ABR,ABRII}

\[
\langle 0 , \widetilde 0 \vert \phi (x^\pm ; \beta ) \phi (0 ; \beta ) \vert \widetilde 0 , 0 \rangle \,=\,D^{(+)} (x^\pm ; \beta  )\,=\,-\,\frac{1}{4 \pi}\,\ln \Big \{i \mu \frac{\beta}{\pi}\,\sinh \frac{\pi}{\beta}(x^\pm - i \epsilon ) \Big \}\,+\,\frac{1}{2\pi} {\it z} (\beta, \mu^\prime)\,,
\]
\[
\langle 0 , \widetilde 0 \vert \widetilde\phi (x^\pm ; \beta ) \widetilde\phi (0 ; \beta ) \vert \widetilde 0 , 0 \rangle \,=\,-\,\frac{1}{4 \pi}\,\ln \Big \{i \mu \frac{\beta}{\pi}\,\sinh \frac{\pi}{\beta}(x^\pm + i \epsilon ) \Big \}\,-\,\frac{i}{4}\,+\,\frac{1}{2\pi} {\it z} (\beta, \mu^\prime)\,,
\]
\[
\langle 0 , \widetilde 0 \vert \eta (x^\pm ; \beta ) \eta (0 ; \beta ) \vert \widetilde 0 , 0 \rangle \,=\,+\,\frac{1}{4 \pi}\,\ln \Big \{i \bar\mu \frac{\beta}{\pi}\,\sinh \frac{\pi}{\beta}(x^\pm - i \epsilon ) \Big \}\,-\,\frac{1}{2\pi} {\it z} (\beta, \bar\mu^\prime)\,,
\]
\[
\langle 0 , \widetilde 0 \vert \widetilde\eta (x^\pm ; \beta ) \widetilde\eta (0 ; \beta ) \vert \widetilde 0 , 0 \rangle \,=\,+\,\frac{1}{4 \pi}\,\ln \Big \{i \bar\mu \frac{\beta}{\pi}\,\sinh \frac{\pi}{\beta}(x^\pm + i \epsilon ) \Big \}\,+\,\frac{i}{4}\,-\,\frac{1}{2\pi} {\it z} (\beta, \bar\mu^\prime)\,.
\]

\no For the off-diagonal two-point functions one has

\[
\langle 0 , \widetilde 0 \vert \widetilde\phi (x^\pm ; \beta ) \phi ( 0 ; \beta ) \vert \widetilde 0 , 0 \rangle\,=\,D^{(+)} (x^\pm\,-\,i\frac{\beta}{2} ; \beta )\,,
\]
\[
\langle 0 , \widetilde 0 \vert \widetilde\eta (x^\pm ; \beta ) \eta ( 0 ; \beta ) \vert \widetilde 0 , 0 \rangle\,=\,-\,D^{(+)} (x^\pm\,-\,i\frac{\beta}{2} ; \beta )\,,
\]

\no where $\mu^\prime$, $\bar\mu^\prime$ are infrared cutoffs, and ${\z} (\beta \mu^\prime)$ is the  mean number of massless particles having momenta in the range $[\mu^\prime , \infty ]$ (the same for ${\z} (\beta \bar\mu^\prime )$)

\be\label{z}
{\z} (\beta \mu^\prime )\,=\,\frac{1}{\pi}\,\int_{\mu^\prime}^\infty \frac{dp}{p} N (p , \beta )\,.
\ee

\no In this way one has for the field combinations $\varphi$ and $\widetilde\varphi$, acting as potentials for the longitudinal currents, the following space- and 
time-independent thermal two-point functions,

\be\label{ctcf1}
\langle 0 (\beta ) \vert \varphi (x) \varphi (y) \vert 0 (\beta ) \rangle\,=\,
\langle 0 (\beta ) \vert \widetilde\varphi (x) \widetilde\varphi (y) \vert 0 (\beta ) \rangle\,=\,
-\,\frac{1}{4 \pi}\,\ln \Big (\frac{\mu}{\bar\mu}\,e^{-\,4\,\big ( \z (\beta \mu^\prime )\,-\,\z (\beta \bar\mu^\prime \big )} \Big )\,,
\ee
\be\label{ctcf2}
\langle 0 (\beta ) \vert \widetilde\varphi (x) \varphi (y) \vert 0 (\beta ) \rangle\,=\,
+\,\frac{1}{4 \pi}\,\ln \Big (\frac{\mu}{\bar\mu}\,e^{-\,4\,\big ( \z (\beta \mu^\prime )\,-\,\z (\beta \bar\mu^\prime \big )} \Big )\,.
\ee

\no Contrary to the zero temperature case, (see eq. (\ref{vphitildevphi})),  at $T \neq 0$ the fields $\varphi$ and $\widetilde\varphi$, exhibit a non-zero, constant, crossed correlation function, as expressed by (\ref{ctcf2}). As we shall see, this will play a fundamental role in the thermal selection rule.
We have again

\[
\langle 0 , \widetilde 0 \vert \lc_\mu (x ; \beta ) \lc_\nu (y ; \beta ) 
\vert \widetilde 0 , 0 \rangle = 0\,,\,\,\,\forall (x , y)\,,
\]
\[
\langle 0 , \widetilde 0 \vert \widetilde{\lc}_\mu (x ; \beta ) 
\widetilde{\lc}_\nu (y ; \beta ) 
\vert \widetilde 0 , 0 \rangle = 0\,,\,\,\,\forall (x , y)\,,
\]
\[
\langle 0 , \widetilde 0 \vert \widetilde{\lc}_\mu (x ; \beta ) {\lc}_\nu (y ; \beta ) 
\vert \widetilde 0 , 0 \rangle = 0\,,\,\,\,\forall (x , y)\,.
\]

\subsection{Thermal Selection Rule}

In order to discuss the thermal selection rule, we shall  consider the general thermal correlation functions of the gauge  invariant field operators $\Psi$ and $\widetilde\Psi$ defined in (\ref{ph}) and (\ref{tph}). Since these fields are given in terms of Wick-ordered exponentials of massive and massless free scalar fields, one should review some aspects of the thermofield bosonization introduced in Refs. \cite{ABR,ABRII}. The Wick-ordered thermofield exponentials of the fields $\phi$, $\eta$ and $\Sigma$ are obtained from the corresponding exponentials at zero temperature as follows  (the same for the corresponding tilde fields)

\[
{\cal U} [-\,\vartheta (\beta ) ]\,\dt\,e^{\displaystyle\,i\,\lambda\,\phi (x^\pm)}\,\dt\,{\cal U} [\vartheta (\beta ) ]\,=\,e^{\displaystyle\,-\,\frac{\lambda^2}{4}\,\z (\beta \mu^\prime )}\,\dt\,e^{\displaystyle\,i\,\lambda\,\phi (x^\pm ; \beta )}\,\dt\,,
\]
\[
{\cal U} [-\,\vartheta (\beta ) ]\,\dt\,e^{\displaystyle\,i\,\lambda\,\eta (x^\pm)}\,\dt\,{\cal U} [\vartheta (\beta ) ]\,=\,e^{\displaystyle\,+\,\frac{\lambda^2}{4}\,\z (\beta \bar\mu^\prime )}\,\dt\,e^{\displaystyle\,i\,\lambda\,\eta (x^\pm ; \beta )}\,\dt\,,
\]
\be\label{Bog-transf}
{\cal U} [-\,\vartheta_\Sigma (\beta ) ]\,\dt\,e^{\displaystyle\,i\,\lambda\,\Sigma (x)}\,\dt\,{\cal U} [\vartheta (\beta ) ]\,=\,e^{\displaystyle\,-\,\frac{\lambda^2}{2}\,\z_{_{\Sigma}} (\beta m )}\,\dt\,e^{\displaystyle\,i\,\lambda\,\Sigma (x ; \beta )}\,\dt\,,
\ee

\no where $\z (\beta \mu^\prime)$ ($\z (\beta \bar\mu^\prime)$) is defined in (\ref{z}), ${\it z}_{_{_\Sigma}} (\beta \, m )$ is the mean number of particles with mass $m$ having
momenta $p^1 \equiv p$ in the range $[0 , \infty ]$,

\be\label{zSigma}
{\it z}_{_\Sigma} (\beta \, m ) = \frac{1}{\pi}\,\int_0^\infty \frac{dp}{\sqrt{p^2 + m^2}}
\,N_\Sigma (p^0  , \beta )\,,
\ee

\no and the  Wick-ordered exponential of the free thermofields are defined by  \cite{ABR,ABRII} 

\[
\dt e^{\displaystyle i\, \varphi(x;\beta)} \dt = e^{\displaystyle i\, \varphi^{(-)}(x;\beta)}e^{\displaystyle i\,\varphi^{(+)}(x;\beta)} \,,
\] 

\no where the ordering is with respect to the {\it zero temperature} creation and annihilation operators. 
Taking this into account, one finds for the gauge invariant thermofields

\[
\Psi_\alpha (x ; \beta)\,=\,{\cal U}[ - \vartheta (\beta ) ] \,\Psi (x)\, {\cal U} [ \vartheta (\beta ) ]\,=\,\Big ( \frac{\overline\mu\,\mZ_\Sigma (\beta  m )}{2 \pi} \Big ) ^{\frac{1}{2}}\,\dt e^{\displaystyle\,i\,\sqrt\pi\,\gamma^5_{\alpha\alpha} \Sigma (x ; \beta )}\dt \sigma_\alpha (x ; \beta )\,,
\]
\[
\widetilde\Psi_\alpha (x ; \beta )\,=\,{\cal U} [ - \vartheta (\beta ) ] \,\widetilde\Psi (x)\, {\cal U} [ \vartheta (\beta ) ]\,=\,\Big ( \frac{\overline\mu\,\mZ_\Sigma (\beta  m )}{2 \pi} \Big ) ^{\frac{1}{2}}\,\dt e^{\displaystyle\,i\,\sqrt\pi\,\gamma^5_{\alpha\alpha} \widetilde\Sigma_\alpha (x ; \beta )}\dt \widetilde\sigma (x ; \beta )\,,
\]

\no where we have defined 

\[
\mZ_{_\Sigma} (\beta \, m) = e^{\displaystyle\,- {\z}_{_{_{\Sigma}}} (\beta  \, m)}\,.
\]

\no The thermal operators $\sigma (x ; \beta )$ and  $\widetilde\sigma (x ; \beta )$, generalizing to finite temperature those introduced in Ref. \cite{LS}, are,

\be\label{sigmaT}
\sigma_\alpha (x ; \beta )\,=\,e^{\displaystyle\,2\,i\,\sqrt\pi\,\varphi (x^\pm ; \beta )}\,=\,\Big (\frac{\mu}{\bar\mu}\,e^{\displaystyle\,-\,2\,( \z( \beta  \mu^\prime )\,-\,\z (\beta \bar\mu^\prime ))} \Big )^{\frac{1}{2}}\,\dt e^{\displaystyle\,2\,i\,\sqrt\pi\,\varphi (x^\pm ; \beta )}\,\dt\,,
\ee
\be\label{sigmaTtilde}
\widetilde\sigma_\alpha (x ; \beta )\,\,=\,e^{\displaystyle\,2\,i\,\sqrt\pi\,\widetilde\varphi (x^\pm ; \beta )}\,=\,\Big (\frac{\mu}{\bar\mu}\,e^{\displaystyle\,-\,2\,( \z( \beta  \mu^\prime )\,-\,\z (\beta \bar\mu^\prime )} \Big )^{\frac{1}{2}}\,
\dt \,e^{\displaystyle\,2\,i\,\sqrt\pi\,\widetilde\varphi (x^\pm ; \beta ) }\dt\,,
\ee

\no  where $x^+$($x^-$) goes with $\sigma_1$($\sigma_2$).  The general thermal correlation function of the gauge invariant field operators $\Psi$ and $\widetilde\Psi$ is then 

\[
\langle 0 (\beta ) \vert \prod_{i = 1}^{n} \Psi (x_i)\,
\prod_{j = 1}^{\tilde n} \widetilde\Psi (\tilde x_j)\,\prod_{k = 1}^{n^\prime} \Psi^\dagger (y_k)\,\prod_{\ell = 1}^{\tilde n^\prime} \widetilde\Psi (\tilde y_\ell) \vert 0 (\beta ) \rangle\,=\,\Big ( \frac{\overline\mu}{2 \pi} {\mZ}_{_\Sigma} (\beta  m) \Big )^{\frac{1}{2}\big ( n^\prime + n + \tilde n^\prime + \tilde n \big )}\times
\]
\[
e^{\textstyle \,-\, \pi \,F_{_\Sigma}(x,y ; \beta)}\,
e^{\textstyle\,-\, \pi H_{_\Sigma} (x , \tilde x, y, \tilde y ; \beta )} \,
\langle 0 , \widetilde 0 \vert \prod_{i = 1}^{n} \sigma_{\alpha_i} (x_i ; \beta )\,
\prod_{j = 1}^{\tilde n} \widetilde\sigma_{\alpha_j} (\tilde x_j ; \beta )\,\prod_{k = 1}^{n^\prime} \sigma^\dagger_{\alpha_k} (y_k ; \beta )\,\prod_{\ell = 1}^{\tilde n^\prime} \widetilde\sigma^\dagger_{\alpha_\ell} (y_\ell ; \beta ) \vert \widetilde 0 , 0 \rangle\,.
\]

\no The functions $F_{_\Sigma}$ and $H_{_\Sigma}$ collect the space-time dependent contributions of the two-point functions of the massive thermofields, $\Sigma$ 
and $\widetilde\Sigma$. These functions, $F_{_\Sigma}$ and $H_\Sigma$, generalize to finite temperature the gauge invariant subspace of the fermion Wightman function at zero temperature obtained by Schwinger \cite{Schwinger,LS} and are given by

\[
F_{_\Sigma}(x , y ; \beta)\,=\,
\sum_{i < i^\prime}^n \gamma^5_{x_i} \gamma^5_{x_{i^\prime}} \Delta ^{(+)} (x_i - x_{i^\prime}  ; m ,\beta ) \,,+
\sum_{k < k^\prime}^{n^\prime} \gamma^5_{y_k} \gamma^5_{y_{k^\prime}} \Delta ^{(+)} (y_k - y_{k^\prime}  ; m,  \beta ) + 
\sum_{i = 1}^n \sum_{k = 1}^{ n^\prime} \gamma^5_{x_i} \gamma^5_{x_{k}} \Delta ^{(+)} (x_i\,-\, y_{k}  ;m , \beta )\,,
\]

\no and 

\[ 
H_{_\Sigma} (x , \tilde x , y , \tilde y ; \beta )\,=\,
\sum_{j < j^\prime}^{\tilde n} \gamma^5_{\tilde x_j} \gamma^5_{\tilde x_{j^\prime}} \Delta ^{(+)} (\tilde x_j - \tilde x_{j^\prime}  ; m, \beta ) +
\sum_{\ell < \ell^\prime}^{\tilde n^\prime} \gamma^5_{\tilde y_\ell} \gamma^5_{\tilde y_{\ell^\prime}} \Delta ^{(+)} (\tilde y_\ell - \tilde y_{\ell^\prime}  ; m,  \beta )
\]
\[
+\,\sum_{i = 1}^n \sum_{j = 1}^{\tilde n} \gamma^5_{x_i} \gamma^5_{\tilde x_{j}} 
\Delta ^{(+)} ({\underline x}_i\, - \tilde x_{j}  ; m,  \beta )
\,-\,
\sum_{i = 1}^n \sum_{\ell = 1}^{\tilde n^\prime} \gamma^5_{x_i} \gamma^5_{\tilde y_{\ell}} \Delta ^{(+)} ({\underline x}_i) - \tilde y_{\ell}  ;m , \beta )
-\,
\sum_{j = 1}^{\tilde n} \sum_{k = 1}^{n^\prime} \gamma^5_{\tilde x_j} \gamma^5_{ y_{k}} \Delta^{(+)} ({\tilde {\underline x}}_j)\, -  y_{k}  ;m , \beta )
$$
\[
+ \sum_{j = 1}^{\tilde n} \sum_{\ell = 1}^{\tilde n^\prime} \gamma^5_{\tilde x_j} \gamma^5_{ \tilde y_{k}} \Delta ^{(+)} (\tilde x_j\, -\,  \tilde y_{\ell}  ; m, \beta )
-\,\sum_{k = 1}^{n^\prime} \sum_{\ell = 1}^{\tilde n^\prime} \gamma^5_{ y_k} \gamma^5_{ \tilde y_{\ell}} \Delta^{(+)} ( y_k\,- \, \tilde {\underline y}_{\ell}  ; m ,  \beta )\,,
\]

\no where (see Appendix )

\[
\langle 0 , \widetilde 0 \vert \Sigma (x ; \beta ) \Sigma (y ; \beta ) \vert \widetilde 0 , 0 \rangle \,=\,
\langle 0 , \widetilde 0 \vert \widetilde\Sigma (x ; \beta ) \widetilde\Sigma (y ; \beta ) \vert \widetilde 0 , 0 \rangle \,=\,\Delta^{(+)} (x\,-\,y ; m , \beta )\,,
\]
\be\label{off2pf}
\langle 0 , \widetilde 0 \vert \widetilde\Sigma (x ; \beta ) \Sigma (\tilde y ; \beta ) \vert \widetilde 0 , 0 \rangle \,= - \,\Delta^{(+)} ({\underline x}\,-\,y ; m , \beta ),
\ee
with
\[
{\underline x} = (x^0 - i{\beta\over 2}, x^1)\,.
\]

\no Note that Eq. (\ref{off2pf}) is in agreement with Umezawa's view of the thermal ``off-diagonal" functions as an analytic continuation of the ``diagonal" ones \cite{U,TFD}, and as we show in the Appendix, is a general structural property of the thermofield dynamics formalism. 

The thermal operators $\sigma (x ; \beta )$ and $\widetilde\sigma (x ; \beta )$ generate constant correlation functions,

$$
\langle 0 , \widetilde 0 \vert \prod_{i = 1}^{n^\prime} \sigma^\dagger_{\alpha_i} (x_i ; \beta )\,
\prod_{j = 1}^{\tilde n^\prime} \widetilde\sigma^\dagger_{\alpha_j} (\tilde x_j ; \beta )\,\prod_{k = 1}^{n} \sigma_{\alpha_k} (y_k ; \beta )\,\prod_{\ell = 1}^{\tilde n} \widetilde\sigma_{\alpha_\ell} (y_\ell ; \beta ) \vert \widetilde 0 , 0 \rangle
$$
\be\label{sigmacf}
\equiv\,
\langle 0 , \widetilde 0 \vert  \sigma_1^{\dagger^{n^\prime_1}} (\beta )\,
\widetilde\sigma_1^{\dagger^{\tilde n^\prime_1}} (\beta )\,\sigma_1^{n_1} (\beta )\, \widetilde\sigma_1^{\tilde n_1} (\beta ) \vert \widetilde 0 , 0 \rangle\,\times 
\langle 0 , \widetilde 0 \vert  \sigma_2^{\dagger^{n^\prime_2}} (\beta )\,
\widetilde\sigma_2^{\dagger^{\tilde n^\prime_2}} (\beta )\,\sigma_2^{n_2} (\beta )\, \widetilde\sigma_2^{\tilde n_2} (\beta ) \vert \widetilde 0 , 0 \rangle\,.
\ee

\no where the expectation value has been factorized into its positive and negative chirality parts, and $\sigma_\alpha(\beta) \equiv \sigma_\alpha(0,\beta), \tilde{\sigma}_\alpha(\beta) \equiv \tilde{\sigma}_\alpha(0,\beta)$.  As we now show, this expectation value implements the charge and chirality
thermal selection rules. Using (\ref{sigmaT}) and (\ref{sigmaTtilde}) we obtain  for each chirality the thermal correlation function 

\[
\langle 0 , \widetilde 0 \vert  \sigma^{\dagger^{n^\prime}} (\beta )\,
\widetilde\sigma^{\dagger^{\tilde n^\prime}} (\beta )\,\sigma^{n} (\beta )\, \widetilde\sigma^{\tilde n} (\beta ) \vert \widetilde 0 , 0 \rangle\,
= \left(\frac{\mu e^{\displaystyle - 2 z(\beta\mu^\prime)}}{\overline\mu e^{\displaystyle - 2 z(\beta\bar\mu^\prime)}}\right)^{\displaystyle\frac{1}{2}\,[(n - n^\prime) - (\tilde n - \tilde n^\prime)]^2}\,.
\]

\no In the limit $\mu \to 0$ the result will be different from zero and independent of the cutoffs $\bar\mu$ and $\bar\mu^\prime$,  if we require that 
\footnote{By considering the free massless scalar thermofield theory as the zero mass limit of the free massive scalar thermofield theory, the regulator $\mu$ of the zero temperature two-point function should be identified with the infrared cutoff $\mu^\prime$ as $\displaystyle \mu^\prime = \frac{\mu}{\pi}$ \cite{ABR,ABRII}. Under this assumption, for a fixed temperature, the infrared singular asymptotic behavior of the integral (\ref{z}) is given by

$$
{\it z} (\beta \,\mu ) _{\mu \approx 0} \rightarrow \frac{1}{\beta \mu}\,+\,\frac{1}{2}\,\ln (\beta \mu)\,.
$$
\no and one has for the two-point function (\ref{ctcf1})
$$
\langle 0 (\beta ) \vert \varphi (x) \varphi(y) \vert  0 (\beta ) \rangle \,=\,-\,\frac{1}{4\pi}\ln{\Big (\frac{e^{\displaystyle\,-\,\frac{2 \pi}{\beta \mu}}}{\displaystyle \beta \overline\mu}\,e^{\displaystyle 2 \,z (\beta\bar\mu^\prime)}\Big )}\,.
$$
\no In this case the selection rule arises from the limit 
$$
\langle 0 , \widetilde 0 \vert  \sigma^{\dagger^{n^\prime}} (\beta )\,
\widetilde\sigma^{\dagger^{\tilde n^\prime}} (\beta )\,\sigma^{n} (\beta )\, \widetilde\sigma^{\tilde n} (\beta ) \vert \widetilde 0 , 0 \rangle\,
= \lim_{\mu \rightarrow 0}\,\Big (\displaystyle \frac{e^{\displaystyle - \frac{2 \pi}{\displaystyle \beta\mu}}}{\beta \overline\mu} \,e^{\displaystyle 2z(\beta\bar\mu^\prime)}\Big )^{\displaystyle \frac{1}{2}\,[(n - n^\prime) - (\tilde n - \tilde n^\prime)]^2}\,=\,\delta_{n - n^\prime , \tilde n - \tilde n^\prime}\,.
$$

\no Without loss of generality, and for economy of arbitrary constants one may choose  $\mu^\prime = \bar\mu^\prime$, such that the temperature-dependent infrared contributions cancel and one finds for (\ref{ctcf1}) the constant correlation functions
$$
\langle 0 (\beta ) \vert \varphi (x) \varphi(y) \vert  0 (\beta ) \rangle \,=\,-\,
\langle 0 (\beta ) \vert \widetilde\varphi (x) \varphi(y) \vert  0 (\beta ) \rangle \,=\,-\,\frac{1}{4\pi}\ln \frac{\mu}{\bar \mu} \,.
$$
\no In this case, one has
$$
\langle 0 , \widetilde 0 \vert  \sigma^{\dagger^{n^\prime}} (\beta )\,
\widetilde\sigma^{\dagger^{\tilde n^\prime}} (\beta )\,\sigma^{n} (\beta )\, \widetilde\sigma^{\tilde n} (\beta ) \vert \widetilde 0 , 0 \rangle\,
= \lim_{\mu \rightarrow 0}\,\Big (\frac{\mu}{\bar\mu} \Big )^{\displaystyle \frac{1}{2}\,[(n - n^\prime) - (\tilde n - \tilde n^\prime)]^2}\,=\,\delta_{n - n^\prime , \tilde n - \tilde n^\prime}\,.
$$
}

\[
n-n' = \tilde n - \tilde n'\,.
\]

\no Hence, for each chirality we obtain the thermal selection rule

\be\label{selection-rule} 
\langle 0 , \widetilde 0 \vert  \sigma^{\dagger^{n^\prime}} (\beta )\,
\widetilde\sigma^{\dagger^{\tilde n^\prime}} (\beta )\,\sigma^{n} (\beta )\, \widetilde\sigma^{\tilde n} (\beta ) \vert \widetilde 0 , 0 \rangle
=\delta_{n-n',\tilde n - \tilde n'}\,.
\ee
We further have
\[
\sigma^\dagger_{\alpha}(\beta) \sigma_{\alpha}(\beta) = 1 =  
\widetilde\sigma^\dagger_{\alpha}(\beta) \widetilde\sigma_{\alpha}(\beta)\,.
\] 

The thermal selection rule (\ref{selection-rule})  has a natural interpretation.
Since the algebraic relations at $T = 0$ are retained at finite temperature, the commutation relations (\ref{a1})-(\ref{a4}) and (\ref{a5}) hold at finite temperature. The operators $\sigma_\alpha(\beta)$ and $\widetilde\sigma_\alpha(\beta )$ carry the  charge and chirality of the free fermion thermofield. However, in contrast to the zero temperature case, the thermal vacuum state $\vert 0 (\beta ) \rangle$ is neither an eigenstate of the free charge operator ${\cal Q}_f$  nor $\widetilde{\cal Q}_f$. Nevertheless, for  the total free charge of the combined system defined by

\[
\widehat{\cal Q}_f\,=\,{\cal Q}_f\,-\,\widetilde{\cal Q}_f\,,
\]

\no one finds
\[
[\widehat{\cal Q}_f ,\sigma_\alpha^{n_\alpha}(\beta)\widetilde{\sigma}_\alpha^{\tilde n_\alpha}(\beta)]\,=\,-\,( n_\alpha\,-\,\tilde n_\alpha )\,\sigma_\alpha^{n_\alpha}(\beta)\widetilde{\sigma}_\alpha^{\tilde n_\alpha}(\beta) \,.
\]

\no The bosonized expression for the zero-temperature Hamiltonians $H$ and $\widetilde H$ are given in terms of the free field Hamiltonians of the Bose fields ($\Sigma, \widetilde\Sigma, \eta, \widetilde\eta, \phi, \widetilde\phi$) as follows

\[
H\,=\,H^{(0)}_{\Sigma}\,+\,H^{(0)}_\eta\,+\,H^{(0)}_\phi\,,
\]
\[
\widetilde H\,=\,\widetilde H^{(0)}_{\tilde\Sigma}\,+\,H^{(0)}_{\tilde\eta}\,+\,H^{(0)}_{\tilde\phi}\,.
\]

\no  The thermal vacuum state $\vert 0 (\beta ) \rangle$ is not an eigenstate of neither $H$ nor $\widetilde H$ \cite{Das}, and thus the thermal states

\[
\vert n_\alpha\,,\,\tilde n_\alpha ; \beta \rangle \,=\,\sigma_\alpha^n\,\widetilde \sigma_\alpha^{\tilde n} \vert 0 (\beta ) \rangle\,,
\]

\no  are not vacuum eigenstates of $H$ or $\widetilde H$. This follows from the fact that, since the $H$ and $\widetilde H$ commute with $\sigma$ and $\widetilde\sigma$, we get

\[
H\,\vert \widetilde n\,,\, n \,; \beta \rangle\,=\,\sigma^n \widetilde\sigma^{\tilde n}\,H\, \vert 0 (\beta ) \rangle\,
=\,\sigma^n \widetilde\sigma^{\tilde n}\,{\cal U} [ \vartheta (\beta ) ] \,H (\beta )\, \vert \widetilde 0 , 0 \rangle\,\neq 0\,,
\]
\[
\widetilde H\,\vert \widetilde n\, ,\, n\, ; \beta \rangle\,=\,\sigma^n \widetilde\sigma^{\tilde n}\,\widetilde H\, \vert 0 (\beta ) \rangle\,=\,
\sigma^n \widetilde\sigma^{\tilde n}\,{\cal U} [\vartheta (\beta ) ]\,\widetilde H (\beta )\, \vert \widetilde 0 , 0 \rangle \,\neq 0\,.
\]

\no However, $\vert n\,,\,\tilde n\, ; \beta \rangle $ are eigenstates  of the total Hamiltonian $\widehat H = H - \widetilde H$ of the combined system  $\widehat H (\beta )\,=\,\widehat H $.
\footnote{This follows from the fact that

$$
{\cal U} [\vartheta (\beta ) ]\,\big ( a^\dagger (p) a (p)\,-\,\tilde a^\dagger (p) \tilde a (p) \big ) {\cal U}^{\,- 1} [ \vartheta (\beta ) ]\,=\, a^\dagger (p) a (p)\,-\,\tilde a^\dagger (p) \tilde a (p)\,.
$$
})
\[
\widehat H\,\vert n_\alpha\,,\, \tilde n_\alpha \,; \beta \rangle\,=\,\sigma^n\,\widetilde\sigma^{\tilde n}\,\widehat H\,\vert 0 ( \beta ) \rangle\,=\,
\sigma^n\,\widetilde\sigma^{\tilde n}\,{\cal U} [ \vartheta (\beta ) ]\,\widehat H\,\vert \widetilde 0 , 0 \rangle\,=\,0\,,
\]
since $\widehat H (\beta )\,=\,\widehat H$. For the total bare charge of the combined system $\widehat{\cal Q}_f$ one finds

\[
\widehat{\cal Q}_f \vert n , \tilde n ; \beta \rangle\,=\,-\,\big ( n\,-\,\tilde n \big )\,\vert n , \tilde n ; \beta \rangle\,.
\]

\no At finite temperature the free fermionic charge and chirality of the total combined system are condensed into the vacua states.

\subsection{Thermal Theta Vacuum States }

For $T = 0$ we have no correlation between the fields $\sigma$ and $\widetilde\sigma$, and hence two independent selection rules for each chirality, corresponding to the conservation of the independent quantum numbers $q_\alpha$ ($q^5_\alpha$) and $\tilde q_\alpha$ ($\tilde q^5_\alpha$) carried by the vacua $\vert n_\alpha \rangle \otimes \vert \tilde n_\alpha \rangle$. However, for $T \neq 0$ we have correlations between fields $\sigma (\beta)$ and $\widetilde\sigma (\beta)$, such that the total Hilbert space of thermal states does not factorizes as a direct product. As we now show only one selection rule emerges in this case for each chirality, corresponding to the conservation of the total quantum number $\hat q_\alpha = q_\alpha - \tilde q_\alpha$ of the combined system. Thus one expects the existence of only one set of theta vacua , say $\{\theta_\alpha \}$, in accordance with the tilde conjugation rule, and  

\[
\langle 0 (\beta ) \vert \sigma_\alpha (x) \widetilde\sigma_\alpha (y ) \vert 0 (\beta ) \rangle\,=\,1\,,
\]
 
\no independent of the vacuum representation. 

Let us  now perform the statistical ensemble average with respect to the thermal $\theta$-vacuum. This is achieved by replacing the zero temperature Fock vacuum $\vert \widetilde 0 , 0 \rangle$ in (\ref{thetavacuum}) by the thermal vacuum state $\vert 0 (\beta ) \rangle$. For notational
simplicity we shall use the notation introduced in  Section 2. One has

\[
\vert \theta \,,\, \widetilde\theta ; \beta \rangle \,\doteq  \, \frac{1}{(2 \pi)^2}\,\sum_{n = - \infty}^{+ \infty}\,e^{\,-\,i\,n\, \theta}\,\sum_{\tilde n = - \infty}^{+ \infty}\,e^{\,\,i\,\tilde n\, \widetilde\theta}\,\sigma^n\, \widetilde\sigma^{\tilde n} \vert 0 (\beta ) \rangle\,=
\,\frac{1}{(2 \pi)^2}\,\sum_{n = - \infty}^{+ \infty}\,e^{\,-\,i\,n\, \theta}\,\sum_{\tilde n = - \infty}^{+ \infty}\,e^{\,\,i\,\tilde n\, \widetilde\theta}\,\vert n , \tilde n ; \beta \rangle\,.
\]

\no We still have the "spurionization", Eq. (\ref{spurionized}) 

\[
\sigma_\alpha \,\vert \theta\,,\,\tilde\theta \,; \beta \rangle\,=\,e^{\,i\,\theta_\alpha}\,\vert \theta\,,\,\tilde\theta \,; \beta \rangle\,,
\quad
\widetilde\sigma_\alpha \,\vert \theta \,,\,
\tilde\theta\,; \beta \rangle\,=\,e^{\,-\,i\,\widetilde\theta_\alpha}\,\vert \theta\,,\,\tilde\theta\, ; \beta \rangle\,.
\]

\no Using the thermal selection rule (\ref{selection-rule}), the orthonormalization condition among the thermal theta vacua  becomes

\[
\langle \beta ; \widetilde\theta^\prime , \theta^\prime \vert \theta , \widetilde\theta ; \beta \rangle\, = \,\frac 1{(2\pi)^4}\sum_{n , n^\prime}\,e^{\,-\,i\,( n\, \theta\,-\,n^\prime \theta^\prime )}\,\sum_{\tilde n, \tilde n^\prime}\,e^{\,i\,(\tilde n\, \widetilde\theta\,-\,\tilde n^\prime \widetilde\theta^\prime)}\,\delta_{n - n^\prime , \tilde n - \tilde n^\prime} \,.
\]

\no Defining $n_\pm = n \pm n^\prime$, $\tilde n_\pm = \tilde n \pm \tilde n^\prime$, the selection rule (\ref{selection-rule}) now reads $\delta_{n_- , \tilde n_-}$, and we get 

$$
\langle \beta ; \widetilde\theta^\prime , \theta^\prime \vert \theta , \widetilde\theta ; \beta \rangle\,=\,\frac{1}{(2 \pi)^2}\,\sum_{n_+} e^{\,-\,\frac{i}{2}\,n_+  ( \theta - \theta^\prime  )}
\sum_{\tilde n_+} e^{\,\frac{i}{2}\,\tilde n_+  ( \widetilde\theta - \widetilde\theta^\prime  )}
\,\sum_{\tilde n_-} \,e^{\,\frac{i}{2}\,\tilde n_-\,( \widetilde\theta\,+\,\widetilde\theta^\prime  )}\,\sum_{n_-} e^{\,-\,\frac{i}{2}\,n_-\,  ( \theta\,+\,\theta^\prime  )}\,\delta_{n_- , \tilde n_-}
$$
\be\label{oc}
=\,\delta^{(2)}  ( \theta\,-\,\theta^\prime  )\,
\delta^{(2)}  ( \widetilde\theta\,-\,\widetilde\theta^\prime  )\,
\sum_{n_- , \tilde n_-} \,e^{\,-\,\frac{i}{2}\, [\,n_-\, ( \theta + \theta^\prime  )\,+\,\tilde n_-\, ( \widetilde\theta\,+\,\widetilde\theta^\prime ) ]}\,\delta_{n_- , \tilde n_-}\,.
\ee

\no Analogous to the $T=0$ computation, the summation over $n_+$ and $\tilde n_+$ implements
\footnote{In the zero temperature case one has $n_+ = 2 n$, $\tilde n_+ = 2 \tilde n$ and  $n_- = \tilde n_- = 0$.} 
Dirac delta function constraints within each theta vacuum. The last summation in (\ref{oc}), obeying the thermal selection rule, owes its origin to the infinite number of states with the same eigenvalues of the total charge $\hat q_\alpha = n_\alpha - \widetilde n_\alpha$ of the combined system,  which have the same orthogonality relations upon thermalization and ensure the symmetry of the total combined system under the tilde conjugation operation. One finds 

\[
\langle \beta ; \widetilde\theta^\prime , \theta^\prime \vert \theta , \widetilde\theta ; \beta \rangle\,\,=\,(2\pi)^2\,\delta^{(2)} (\theta-\tilde\theta)\,\big [\,\delta^{(2)} (\tilde\theta-\tilde\theta^\prime)\,\delta ^{(2)} \left(\theta-\theta^\prime\right) \big ]\,.
\]

\no In this way we are naturally lead to define the averaged thermal expectation values of observables with respect to the thermal theta vacua states as follows,

\[
\overline{\langle 0( \beta ) \vert\,{\cal P}\,\vert 0 (\beta ) \rangle}\,\doteq\,\frac{1}{(2 \pi)^2}
\int_0^{2\pi} d ^2 \widetilde\theta\,\int_0^{2\pi} d^2 \theta^\prime  \,\int_0^{2\pi}  d^2 \widetilde\theta^\prime\,\,\langle \beta ; \theta^\prime , \widetilde\theta^\prime \vert {\cal P} \vert \widetilde\theta , \theta ; \beta \rangle\,.
\]

\no This result can be interpreted as saying that, within the thermofield dynamics, the thermalization selects just one theta vacuum parameter for each 
chirality out of the two that label the zero temperature theta vacua of the doubled system. This comes about as an 
intrinsic property of the model for the thermal vacuum states, which enable the existence of correlations only within the same thermal theta world. We have

\[
{\overline{\langle 0( \beta ) \vert\,\sigma_\alpha (\beta)\,\vert 0 (\beta ) \rangle}} \,=\,e^{\,i\,\theta_\alpha}\,\,\,\,,\,\,\,
{\overline{\langle 0( \beta ) \vert\,\widetilde\sigma_\alpha (\beta)\,\vert 0 (\beta ) \rangle} \,=\,e^{\,-\,i\,\theta_\alpha}}\,.
\]

\no For a given theta vacuum sector , $\sigma_\alpha$ and $\widetilde\sigma_\alpha$ are $c$-numbers

\[
\sigma_{\alpha }(\beta) = e^{\,i\,\theta_\alpha}\,\,\,\,,\,\,\,
\widetilde\sigma_{\alpha }(\beta) = e^{\,-\,i\,\theta_\alpha} \,,
\]

\no such that

\[
\sigma_{\alpha }(\beta)\,\widetilde\sigma_{\alpha }(\beta)  \equiv 1\,.
\]

\section{Chiral Condensate at Finite Temperature and Symmetry Breaking}
\setcounter{equation}{0}

As is well known, some properties of a two-dimensional quantum field model are more transparent in the operator formulation, and others are better seen in the functional integral formulation. In order to emphasize the usefulness of the thermofield bosonization  we shall compute the expression for the chiral condensate and the corresponding high-temperature behavior. This  enables one to obtain a statistical-mechanical interpretation for the high temperature behavior of the chiral condensate, which is easily obtained within the thermofield bosonization approach. 

In Refs. \cite{Kao,Sachs,AARII}, using the functional integral formulation in Euclidean space, the high-temperature dependence of the chiral condensate in the Schwinger model has been shown to be given by \cite{Sachs}

\[
\langle \bar\psi \psi \rangle_\beta \,\sim\,\frac{1}{\beta} \, e^{\,-\,\frac{\pi}{\beta m}}\,\cos \theta.
\]

\no The dependence on the  $\theta$ vacuum parameter has been considered in Ref. \cite{Kao} as an {\it ad hoc} assumption.

To begin with, let us consider the computation of the theta-vacuum expectation value of the chiral density 
$\overline{\langle 0 (\beta )\vert   \dt \overline{\Psi} (x)  \Psi (x)\dt \vert  0 (\beta ) \rangle}$. To this end we shall consider the mass operator defined by the point-split operator product at zero temperature. From (\ref{ph}) we have

\[
\dt \overline{\Psi} (x)  \Psi (x)\dt =\, \lim_{\epsilon \rightarrow 0}\,
f^{- 1 } \epsilon )\,\overline{\Psi} (x + \epsilon) \Psi (x)\,=\,\frac{\overline\mu}{2 \pi}\,\Big ( \sigma^\dagger_1  \sigma_2 \,\dt e^{\,2\,i\,\sqrt \pi \Sigma (x )}\,\dt \,+\,\sigma^\dagger_2  \sigma_1 \,\dt e^{\,-\,2\,i\,\sqrt \pi \Sigma (x)}\,\dt\Big )\,,
\]

\no where the renormalization constant $f (\epsilon)$ is defined by

\[
f (\epsilon)\,=\,e^{\,-\,\pi \langle 0  \vert \Sigma (\epsilon ) \Sigma (0) \vert  0 \rangle}\,.
\]

\no Now, using (\ref{Bog-transf}) and ({\ref{zSigma}) \cite{ABR,ABRII}, we get

$$
\dt \overline{\Psi} (x , \beta )  \Psi (x , \beta )\dt\,=\, {\cal U} [\vartheta (\beta)] \,\dt \overline{\Psi} (x)  \Psi (x)\dt {\cal U}^{-1} [\vartheta (\beta) ]
$$
\[
=\,
\frac{\overline \mu}{2 \pi}\,e^{\,-\,2\,\pi\,z_{_{\Sigma}} ( \beta m)}\,\Big ( \sigma^\dagger_{1}(\beta) \sigma_{2}(\beta)\,\dt e^{\,2\,i\,\sqrt \pi \Sigma (x ; \beta )}\,\dt \,+\,\sigma^\dagger_{2}(\beta) \sigma_{1}(\beta)\,\dt e^{\,-\,2\,i\,\sqrt \pi \Sigma (x ; \beta )}\,\dt\Big )\,.
\]

\no With respect to the thermal theta vacuum, one finds for the gauge-invariant expression of spontaneous chiral symmetry breaking

\be\label{ccond}
\overline{\langle 0 (\beta ) \vert \dt \overline{\Psi} (x)  \Psi (x)\dt \vert  0 (\beta ) \rangle} \, =\,
\frac{\overline\mu}{\pi}\,e^{\displaystyle\,-\,2\,\pi\, z_{_{\Sigma}} (\beta m)} \,\cos \big ( \theta _2\,-\,\theta_1 \big )\,.
\ee

\no At zero temperature, the chiral condensate behaves as

\[
\overline{\langle 0  \vert \dt \overline{\Psi} (x)  \Psi (x)\dt \vert  0  \rangle} \, =\,
\frac{\overline\mu}{\pi}\,\cos \big ( \theta _2\,-\,\theta_1 \big )\,.
\]

\no From the statistical-mechanical point of view one sees that the thermal chiral condensate is given in terms of the mean number of massive particles in the ensemble (see Appendix). 

The chiral symmetry breaking at finite temperature emerges in accordance with the violation of the cluster decomposition property for the mass operator, which imply the existence of degenerate thermal vacua states. One finds

\be\label{cdp}
\langle 0 (\beta ) \vert \big (\dt \overline{\Psi} (x + \lambda )  \Psi (x + \lambda )\dt \big )\,\big ( \dt \overline{\Psi} (x)  \Psi (x)\dt \big ) \vert 0 (\beta ) \rangle \vert_{\lambda \rightarrow \infty} \rightarrow \,\Big ( \frac{\bar\mu}{ \pi} \Big )^2\,e^{\displaystyle\,-\,4 \pi {\it z}_{_{\Sigma}} (\beta m)}\,\neq\,\big \vert \langle 0 (\beta ) \vert \dt \overline{\Psi} (x)  \Psi (x)\dt \vert 0 (\beta ) \rangle \big \vert^2\,,
\ee

\no since the thermal selection rule (\ref{selection-rule}) would imply

\[
\langle 0 (\beta ) \vert \dt \overline{\Psi} (x)  \Psi (x)\dt \vert 0 (\beta ) \rangle\,=\,0\,.
\]

In order to show that the chiral symmetry remains broken at all finite temperature, let us consider the high-temperature behavior of the chiral condensate. To this end let us consider the  mean number of massive particles at temperature $\beta$ (see Appendix)

\[
{\it z}_{_{\Sigma}} (\beta m)\,=\,\frac{1}{\pi}\int_0^\infty \frac{dp}{\sqrt{p^2 + m^2}}\,\frac{1}{e^{\beta \sqrt{p^2 + m^2}}\,-\,1}
\]
\be\label{Z}
=\,\frac{1}{\pi}\,\sum_{k = 1}^\infty K_0 (k \beta m )\,=\,\frac{1}{2\pi}\, \ln \big ( \frac{m e^\gamma \beta}{4 \pi} \big )\,+\,\frac{1}{2 \beta m}\,+\,\sum_{\ell = 1}^\infty \Big \{ \frac{1}{\sqrt{(\beta m)^2 + (2 \ell \pi)^2}}\,-\,\frac{1}{2 \ell \pi }\Big \}\,.
\ee

\no For high temperatures  the leading terms are 

\be\label{htb}
{\it z}_{_{\Sigma}} (\beta m) \approx\,
\frac{1}{2\pi}\, \ln \big ( \frac{m e^\gamma \beta}{4 \pi} \big )\,+\,\frac{1}{2 \beta m}\,.
\ee

\no From (\ref{htb}) and (\ref{cdp}) one has for high temperatures

\[
\langle 0 (\beta ) \vert \big (\dt \overline{\Psi} (x + \lambda )  \Psi (x + \lambda )\dt \big )\,\big ( \dt \overline{\Psi} (x)  \Psi (x)\dt \big ) \vert 0 (\beta ) \rangle \vert_{\lambda \rightarrow \infty} \approx\,
\Big (\frac{4 \bar\mu}{m e^\gamma} \Big )^2\,\big (  k T \big )^2\,e^{\displaystyle\,-\,\frac{2 \pi }{m}\,kT}\,,
\]

\no which shows that, as expected,  there is no a critical temperature for which the cluster decomposition property is restored, the chiral symmetry remaining broken for all finite temperature. With respect to the $\theta$ vacuum, the high temperature behavior of the chiral condensate (\ref{ccond}) is then  given by

\be
\overline{\langle 0 (\beta )\vert  \dt \overline{\Psi} (x) \Psi (x) \dt \vert  0 (\beta ) \rangle}\,\approx\,\Big (\frac{4 \bar\mu}{m e^\gamma} \Big )\,\big (  k T \big )\,e^{\displaystyle\,-\,\frac{\pi }{m}\,kT}\,\cos \big ( \theta_2\,-\,\theta_1 \big )\,.
\ee

\section{Concluding Remarks}

We have considered the operator solution for Quantum Electrodynamics in two dimensions at finite temperature within the thermofield dynamics approach. Using the bosonization scheme developed in refs. \cite{ABR,ABRII} we have obtained the generalization  of the operator solution due to Lowenstein and Swieca \cite{LS} at zero temperature to finite temperature. Our thermofield bosonization approach for the free fermion field is in agreement with the revised version of thermofield dynamics formalism for fermions due to Ojima \cite{O}, which plays a crucial role in obtaining consistency with the free fermion thermofield selection rule.

With respect to the physical content of the model a remark is in order: We have defined as physical states the set of gauge invariant states, regardless of  whether they are tilde or untilde states. From the observable point of view, the fields that describe the physical system under consideration are the untilded gauge invariant operators obtained from $\{\bar\psi,\psi, {\cal A}_\mu\}$. The statistical-mechanical thermal averages of these operators are those of physical relevance. However, the gauge invariant crossed thermal correlation functions (off-diagonal) involving both tilde and non-tilde fields participate as intermediary (virtual) processes in a perturbative computation (as for instance, in the perturbative expansion in the fermion mass for the case of the massive Thirring model \cite{ABRII}).

Within the operator approach one easily shows that the chiral symmetry breaking at $T=0$ persists for any finite temperature; the demonstration of this proved much simpler in this formalism as compared to the functional approach \cite{AARII}. The existence of only one set of theta vacua characterizing the chiral symmetry breakdown emerges here as a byproduct of the thermal selection rule which ensures that the bare fermionic  charge and chirality of the total combined system are screened.

Since the equal-time properties of commutators remain unaffected by the unitary transformation taking one from $T = 0$ to $T \neq 0$, it is furthermore easy to see, that the usual arguments demonstrating the screening of the charge and confinement remain valid.

The authors dedicate this work to the memory of  Jorge Andr\'e Swieca.

 {\bf Acknowledgments}

{ The authors are grateful to the Brazilian Research Council (CNPq) and to DAAD scientific exchange program, which make this collaboration possible. One of us (K.D.R.) wishes to thank the Department of  Physics of the Universidade Federal Fluminense (Brazil) for the kind hospitality extended to him.}

\appendix{{\centerline{\bf{Appendix }}}}
\centerline{\bf Two-Dimensional Free Massive Scalar Thermofield Theory}
\renewcommand{\theequation}{{A}.\arabic{equation}}\setcounter{equation}{0}

The free massive scalar thermofields in two-dimensions are given by \cite{ABR} ($ p^0 = \sqrt{ (p^1)^2 + m^2}$)

\[
\Sigma (x ; \beta ) = {\cal U}^{-1} [\vartheta (\beta ) ] \Sigma (x) {\cal U} [\vartheta (\beta ) ] =
\,\frac{1}{2 \sqrt \pi} \int_{- \infty}^{+ \infty} \frac{dp^1}{\sqrt{p^0}}\,\big \{ e^{\,- \,i p^\mu x_\mu} \big [ a (p^1) \cosh \vartheta ( p^0 ,\beta )\,-\,\tilde a^\dagger (p^1) \sinh \vartheta (p^0 , \beta ) \big ]
\]

\[
+\,e^{\,i\,p^\mu x_\mu} \big [ a^\dagger (p^1) \cosh \vartheta ( p^0 ,\beta )\,-\,\tilde a (p^1) \sinh \vartheta (p^0 , \beta ) \big ] \big \}\,,
\]

\[
\widetilde\Sigma (x ; \beta ) = {\cal U}^{-1} [\vartheta (\beta ) ] \widetilde\Sigma (x) {\cal U} [\vartheta (\beta ) ] =
\,\frac{1}{2 \sqrt \pi} \int_{- \infty}^{+ \infty} \frac{dp^1}{\sqrt{p^0}}\,\big \{ e^{\, \,i p^\mu x_\mu} \big [ \tilde a (p^1) \cosh \vartheta ( p^0 ,\beta )\,-\,a^\dagger (p^1) \sinh \vartheta (p^0 , \beta ) \big ]
\]

\[
+\,e^{\,-\,i\,p^\mu x_\mu} \big [ \tilde a^\dagger (p^1) \cosh \vartheta ( p^0 ,\beta )\,-\, a (p^1) \sinh \vartheta (p^0 , \beta ) \big ] \big \}\,.
\]

\no The diagonal and off-diagonal two-point functions are ($p = \vert p^1 \vert $)

\be\label{Diag}
\langle 0 , \widetilde 0 \vert \Sigma (x ; \beta ) \Sigma (y ; \beta ) \vert \widetilde 0 , 0 \rangle\,=\,\nabla^{(+)} (x - y ; \beta , m)
\ee
\[
= \,\frac{1}{4 \pi}\,\int_0^\infty \frac{d p}{\sqrt{p^2 + m^2}}\Big \{ e^{\,-\,i\,p^\mu (x - y)_\mu}\,\cosh^2 \vartheta (p , \beta )\,+\,e^{\,i\,p^\mu (x - y)_\mu}\,\sinh^2 \vartheta (p , \beta )\Big \}
\]

\be\label{Off}
\langle 0 , \widetilde 0 \vert \widetilde\Sigma (x ; \beta ) \Sigma (y ; \beta ) \vert \widetilde 0 , 0 \rangle\,=\,\nabla^{(+)} ({ x} - y ; \beta , m)
\ee

\[ 
=\,-\,\frac{1}{4 \pi}\,\int_0^\infty \frac{d p}{\sqrt{p^2 + m^2}}\Big \{ e^{\,-\,i\,p^\mu (x - y)_\mu}\,\cosh \vartheta (p , \beta ) \sinh \vartheta (p , \beta )\,+\,e^{\,i\,p^\mu (x - y)_\mu}\,\sinh \vartheta (p , \beta ) \cosh \vartheta (p , \beta )\,\Big \}
\]

\no In order to show that the off-diagonal two-point function (\ref{Off}) corresponds to an analytic continuation of the diagonal ones (\ref{Diag}), we shall use that 
\footnote{This argument is general, independent of the space-time dimensions and also holds for the two-point function of free Fermi thermofields, upon replacing the hyperbolic functions by the corresponding trigonometric functions \cite{Das,U,O},

\[
\sin \vartheta_F ( p^0 , \beta )\,=\,e^{\,-\,\frac{\beta}{2} p^0} \cos \vartheta_F ( p^0 , \beta)\,,
\]
\no with the Fermi-Dirac statistical weight given by

\[
\sin^2 \vartheta_F (p^0 , \beta )\,=\,\frac{1}{e^{\beta p^0} + 1}\,.
\]}
\[
\sinh \vartheta (p^0 , \beta )\,=\,e^{\,-\,\frac{\beta}{2} p^0}\,\cosh \vartheta ( p^0 , \beta )  \,.
\]

\no The off-diagonal two-point function (\ref{Off}) can be written as

\[
\nabla^{(+)} (x - y ; \beta , m)\,
=\,-\,\frac{1}{4 \pi}\,\int_0^\infty \frac{d p}{\sqrt{p^2 + m^2}}\Big \{ e^{\,-\,i\,[ p^\mu (x - y)_\mu\,-\,i\,\frac{\beta}{2} p^0]}\,\cosh^2 \vartheta (p , \beta )\,+\,e^{\,i\,[ p^\mu (x - y)_\mu\,-\,i\,\frac{\beta}{2} p^0 ]}\,\sinh^2 \vartheta (p , \beta )\Big \}\,,
\]

\no that is,

\[
\nabla^{(+)} (x - y ; \beta , m)\,=\,-\,\Delta^{(+)} (\underline{x} - y ; \beta , m )\,,
\]

\no with $\underline{x}^\mu = (x^0 - i \frac{\beta}{2} , x^1 )$.  

Let us now compute the 2-point function (\ref{Diag}), which can be decomposed as

\[
\Delta ^{(+)} (x ; m , \beta ) \,=\,\Delta ^{(+)}_o ( x ; m )\,+\,I (x ; m , \beta )\,,
\]

\no where $\Delta_o^{(+)}$ is the zero temperature contribution 

\[
\Delta ^{(+)}_o ( x ; m )\,=\,\frac{1}{2 \pi} K_0 \big ( m \sqrt {- x^2 } \big )\,,
\]

\no and the finite temperature contribution is given by

\be\label{I1}
I (x ; m , \beta )\,=\,\frac{1}{\pi}\,\int_0^\infty \frac{dp}{\sqrt{p^2 + m^2}}\,\cos \big [ x^0 \sqrt{p^2 + m^2} \big ]\,\cos \big [ p x^1 \big ]\,\sinh^2 \vartheta (p , \beta )\,.
\ee

\no The mean number of particles with mass $m$ corresponds to

\[
{\it z}_{_{\Sigma}} (\beta m) \,=\,I (0 ; \beta m)\,.
\]

\no The Bose-Einstein statistical weight can be written as

\[
\sinh ^2 \theta = \,\frac{1}{e^{\beta \sqrt{p^2 + m^2}} \,-\,1}\,=\,
\frac{1}{2}\,e^{\,-\,\frac{\beta}{2} \sqrt{p^2 + m^2}}\,\mbox{cosech} \big [ \frac{\beta}{2} \sqrt{p^2 + m^2} \big ]\,=\,
\sum_{n = 1}^\infty e^{\,-\,n \beta \sqrt{p^2 + m^2}}\,.
\]

\no The temperature dependent contribution (\ref{I1}) is then

\be\label{I2}
I (x ; \beta , m) = \frac{1}{2 \pi}\,\sum_{n = 1}^\infty\,\int_0^\infty \frac{dp}{\sqrt{p^2 + m^2}}\,\cos (p x^1 )\,\Big ( e^{\,-\,[\, n \beta\,-\,i\,x^0 ] \sqrt{p^2 + m^2}}\,+\,e^{\,-\, [\,n  \beta\,+\,i\,x^0 ] \sqrt{p^2 + m^2}}\,\Big )\,.
\ee

\no Introducing $x^\pm = x^0 \pm x^1$, one can write (\ref{I2}) in terms of a Bessel function series as \cite{Grads}

\[ 
I (x ; \beta , m )\,=\,\frac{1}{2 \pi}\,\sum_{n = 1}^\infty\,\Big \{ K_0 \Big ( m \sqrt{ - (x^+ + i n \beta ) ( x^- + i n \beta )}\Big )\,+\, K_0 \Big ( m \sqrt{ - (x^+ - i n \beta ) (x^- - i n \beta ) } \Big )\Big \}\,.
\]

\no {\bf i) Zero mass limit:}  \\

\no For $m \approx 0$, one has ($\mu = m\,e^\gamma / 2$ and $\gamma$ is de Euler constant)

\[
\Delta_0^{(+)} (x , m) \approx\,-\,\frac{1}{4\pi}\,\ln \big ( \mu^2\,i x^+\,i x^- \big )\,.
\]

\no and for the temperature dependent contribution we have

\[
I ( x ; \beta , m \approx 0 ) \approx\,-\,\frac{1}{4 \pi}\,\sum_{n = 1}^\infty \Big \{ \ln \Big ( (i \mu )^2 (x^+ + i n \beta ) (x^+ - i n \beta ) \Big )
+ \ln \Big ( (i \mu )^2 (x^- + i n \beta ) (x^- - i n \beta ) \Big ) \Big \}\,.
\]

\no Using that \cite{Grads} ($x \equiv \frac{x^\pm \pi}{i \beta}$)

\[
\sum_{n = 1}^\infty\,\ln \Big ( 1\,-\,\frac{x^2}{n^2 \pi^2} \Big )\,=\,\ln ( \sin x )\,-\,\ln x\,\,\,,\,\,\,0 < x < \pi\,,
\]

\no the zero mass limit of the massive thermofield two-point function is given by \cite{ABR}

\[
\Delta^{(+)} (x ; \beta , m \approx 0 ) \approx\,-\,\frac{1}{4 \pi}\,\ln \Big \{ \big [ \frac{i \mu \beta}{\pi} ]^2\,\sinh \big ( \frac{\pi}{\beta} x^+ \big ) \, \sinh \big ( \frac{\pi}{\beta} x^- \big ) \Big \} \, -\,\frac{1}{\pi} \sum_{n = 1}^\infty \ln \big ( i \mu n \beta)\, ,
\]

\no where the last term corresponds to the mean number of massless particles with momentum in the range $[ \mu , \infty ]$. \\

\no {\bf ii) Mean number of massive particles :}\\

\no For the mean number of massive particles in the ensemble, one has 

\[
{\it z}_{_{\Sigma}} (\beta m )\,=\,I (0 ; \beta m)\,=\,\frac{1}{\pi}\,\sum_{n = 1}^\infty K_0 \big (n \beta m \big )\,.
\]

\no Using that \cite{Grads}

\[
\sum_{k = 1}^\infty K_0 (k x ) \cos (k x t)\,=\,\frac{1}{2} \big ( \gamma \,+\,\ln \frac{x}{4 \pi} \big )\,+\,\frac{\pi}{2 x \sqrt{1 + t^2}}\,
\]
\[
 +\,\frac{\pi}{2}
\,\sum_{\ell = 1}^\infty \Big \{ \frac{1}{\sqrt{x^2 + (2 \ell \pi - t x )^2}}\,-\,\frac{1}{2 \ell \pi} \Big \}\,+\,\frac{\pi}{2}\,\sum_{\ell = 1}^\infty \Big \{ \frac{1}{\sqrt{x^2 + (2 \ell \pi + t x )^2}}\,-\,\frac{1}{2 \ell \pi} \Big \}\,,
\]

\no one obtains the expression (\ref{Z}).

\newpage

\end{document}